\documentclass[prx,twocolumn,showpacs,superscriptaddress]{revtex4-1}
\usepackage[english]{babel}
\usepackage[dvips]{graphicx}
\usepackage{color}
\usepackage{latexsym}
\usepackage{amsmath}
\usepackage{amsthm}
\usepackage{amsfonts}
\usepackage{amssymb}
\usepackage{mathtools}
\usepackage{array}
\usepackage{cancel}
\usepackage{algorithmic}
\usepackage[dvipsnames]{xcolor}
\usepackage[miktex]{gnuplottex}
\usepackage{epstopdf}
\usepackage{bbold}
\begin{document}

\pagenumbering{arabic}

\title{Performance of quantum registers in diamond in the presence of spin impurities}

\author{Dominik Maile}
\author{Joachim Ankerhold}
\affiliation{Institute for Complex Quantum Systems and IQST, Ulm University, D-89069 Ulm, Germany}
\begin{abstract}
 The Nitrogen Vacancy Center in diamond coupled to addressable surrounding nuclear spins forms a versatile building block for future quantum technologies. While previous activities focused on sensing with only a single or very few spins in operation, recently multi-qubit registers have been successfully implemented for quantum information processing. Further progress requires a detailed understanding of the performance of quantum protocols for consecutive gate operations and thus, beyond established treatments for relaxation and dephasing.
 Here, we provide such a theoretical analysis for a small spin registers with up to four spins built out of NV and environmental constituents in presence of ensembles of interacting impurity spins. Adapting a cluster correlation expansion, we predict coherence properties as well as fidelities for GHZ- and Bell-gate operations also in presence of decoupling pulses. The influence of the volume density and the geometry of the spin-bath consisting of the substitutional nitrogen- or $^{13}C$ atoms are also taken into account.
\end{abstract}
\date{\today}
\maketitle

\section{Introduction}
Decoherence and relaxation are prime obstacles for the usability of quantum mechanical attributes of a physical system.
Over the last decades, theoretical and experimental physicists proposed a zoo of different platforms to overcome the fragility of information encoded in quantum superposition states. 
Although several small quantum processors have been built from different platforms - superconducting Josephson arrays and arrays of trapped ions being the most prominent ones - each of them is subject to different limitations with respect to scalability and coherence times.   
In this respect, the Nitrogen Vacancy center (NV) in diamond is in principle a promising  building block for  quantum technologies offering long coherence times and efficient optical readout even at ambient conditions \cite{DOHERTY2013,Mejer2021,Wolfowicz2021}. 
While coherent processes with a single NV-center were already realized almost two decades ago \cite{Jelezko2004-0} and are now routinely exploited for sensing purposes, scaling up to larger NV-registers remains a challenge \cite{Dolde2013,Shinobu2019}. 
To be able to coherently couple individual NV-centers via their dipole interaction efficiently, they have to be located at a distance of a few tens of nanometers \cite{Jakobi_2016}.
Since this distance dependence cannot be controlled with sufficient precision yet, the assembling of larger registers relies on stochastic processes during the diamond growth or implementation of nitrogen \cite{Groot_Berning2021,Shinobu2019,Jakobi_2016,Yamamoto2013}.
%
%
%
%%%%%%%%%%%%%%%%
\begin{figure}[t]
\includegraphics[scale=0.325]{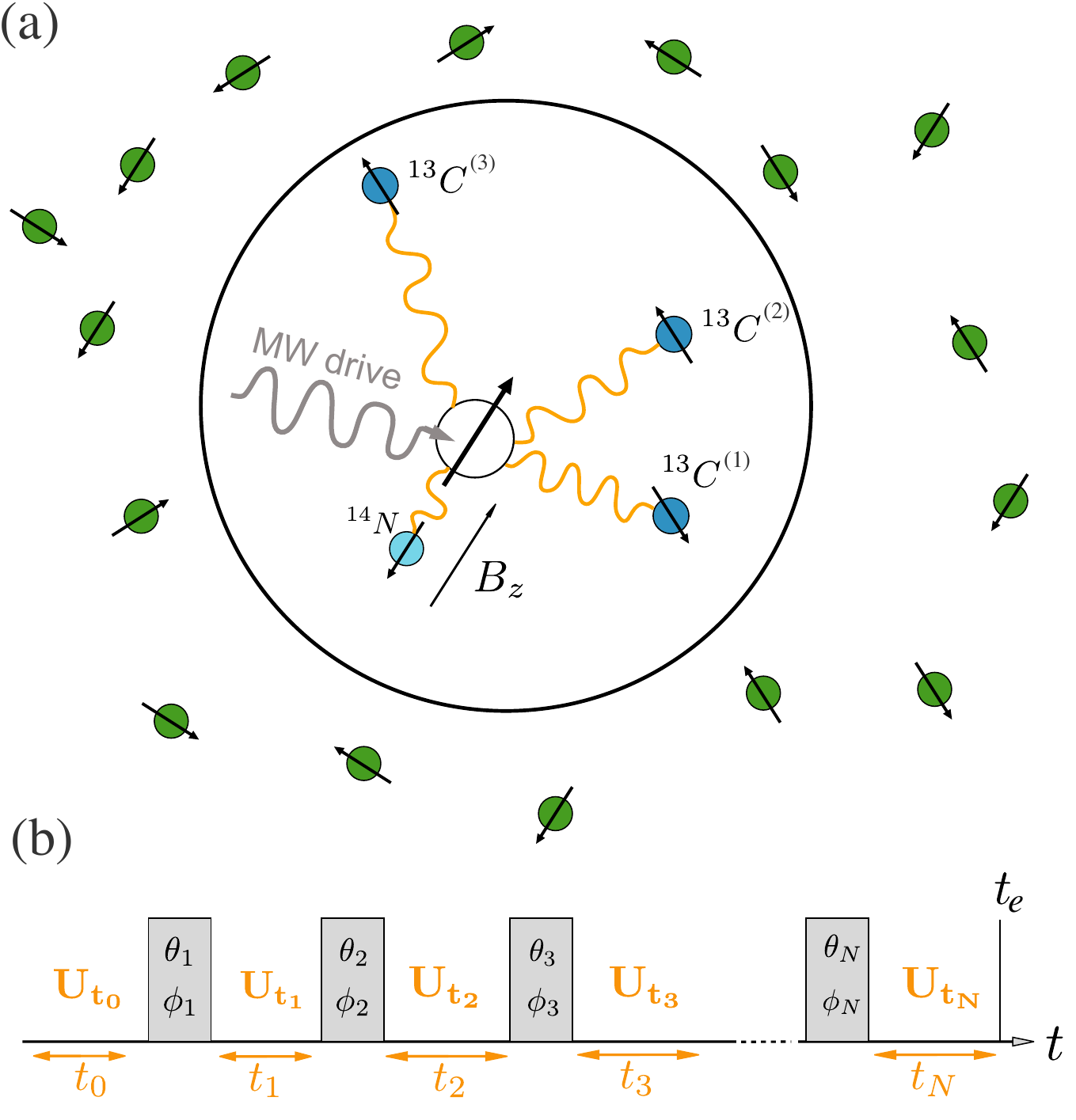}
%\scalebox{.8}{\input{Bild.tex}}
\caption{(a) NV-based setting of a quantum register studied in this work. The bare register (encircled) consists of one $\text{NV}^{-}$ (with one electron spin and one nitrogen nuclear spin) and three $^{13}C$ atoms. The $\text{NV}^{-}$ electron spin is located in the center of the coordinate system and the coordinates of the $^{13}C$ spins are $(0.87,0,0.19)$ nm, $(0.56,0.77,0.31)$ nm and $(-0.83,1.15,0.46)$ nm.  Spins outside of the circle represent the spin-bath consisting of residual $ ^{13}C$-spins or nitrogen-impurities (P1). (b) General form of the pulse structure used to initiate operations on the register. We assume each pulse to be an instantaneous rotation on the Bloch sphere of the $\text{NV}^{-}$ given by the angles $\theta_i,\phi_i$. Between pulses, register+spin-bath evolve unitarily for times $t_i$. Parameters $\theta_i,\phi_i,t_i$ are found by optimizing the fidelity with respect to a specific target state. }
\label{Fig.1}
\end{figure}
%%%%%%%%%%%%%%%%
%
Given this situation, the focus on building NV-based registers has been laid on the coupling to other magnetic impurities found in diamond,
predominantly carbon $^{13}C$ isotopes.

However, also other impurities play an important role for NV based quantum technologies.  Substitutional nitrogen atoms (P1-impurities) are a necessary ingredient to obtain NV-centers in the first place, either during the diamond growth process or by implantation \cite{Meijer2019}. 
Further, there is evidence, that nitrogen is also an electron donor turning the uncharged $\text{NV}^{(0)}$ to the technologically useful negatively charged $\text{NV}^{-}$ making close P1-impurities unavoidable \cite{DOHERTY2013}.
Overall, the creation yield of these nitrogen impurities to pair with vacancies and to become NV-centers, plays a crucial role. Too many remaining pure nitrogen impurities form a spin-bath and  lead to fast decoherence for the register as the surrounding electron spins couple strongly to the NV-center \cite{Bauch2020}.  
As mentioned above, additional $^{13}C$ atoms also have a two-sided influence on the performance of a diamond based spin register: Depending on their abundance and individual position, they may act as a spin-bath or as addressable qubits.  
In fact, the $^{13}C$s which are strongly coupled to the $\text{NV}^{-}$ can be initialized to the desired state and are addressable via either radio-frequency drives and/or via the dipole interaction with the $\text{NV}^{-}$ \cite{Jelezko2004,Childress2006,Neumann2008,Jacques2009}.
The long coherence properties of the nuclear spins also lead to the proposal to use them as quantum memory or quantum repeaters for quantum information networks \cite{Bernien2013,Hanson2015,Taminiau2021,Hanson2021network}. 

Based on the idea of using several $^{13}C$s as being part of the register, quantum gates and small quantum algorithms have successfully been implemented experimentally \cite{vanderSar2012,Suter2020,Suter2020-2,Taminiau2014,Taminiau2019,Taminiau2022-2}.  
These algorithms can be initiated by local control of the full register: MW-driving the $\text{NV}^{-}$ and additional radio-frequency drives for the $^{13}C$.

Most spectacularly, the coherent control of a quantum register consisting of a single NV-center and nine nuclear spins (including the nuclear spin of the NV) was recently realized in such a way \cite{Taminiau2019}.
On the other hand, a minimal control scheme uses only the MW-drive on the $\text{NV}^{-}$ to perform processes on the full register. Here, also one-qubit gates on the $^{13}C$s are mediated by the dipole interaction with the $\text{NV}^{-}$ \cite{Plenio2017,Taminiau2014,Suter2020,Suter2020-2}. 

The main infidelity of NV-registers for processes on the timescale of several microseconds and especially at lower temperature can be attributed to the surrounding spin-bath \cite{Bar-Gill2012,Bar-Gill2013}, where details strongly depend on
the abundance and the specific distribution of impurity spins.
Even by including dynamical decoupling pulses one problem is the growth of unwanted entanglement to residual nuclear spins during the gate sequences. 
Further progress towards quantum computing architectures thus necessitates a detailed understanding of decoherence induced by these environmental degrees of freedom beyond mean field approximations and their impact on gate operations beyond single-qubit processes. 

From the theory side, a well-explored settings consists of a central spin embedded in a spin-bath \cite{DasSarma2003,Sham2006,Lukin2008,Witzel2006,Breuer2004,Hanson2008,LiuNV,LiuReview,Witzel2012,Hall2014,Myers2017,Galli2021,Kwiatkowski2018,Bragar2015,Ivady2020,Grasser2021} or coupled to phonons \cite{Norembuena2018,Norambuena2021,Norambuena2023}. Thereby, the effect of the spin-bath has mainly been considered as a pure dephasing processes only by starting initially from a dedicated quantum state. 
Here, a microscopic approach that takes a specific configuration of the spin-bath into account, uses a "cluster correlation expansion" (CCE) to subsequently include higher correlations within the bath \cite{Witzel2006,LiuNV}. 
This technique was recently generalized to the "generalized cluster correlation expansion" (gCCE) which can be used to calculate the full density matrix of a system also capturing relaxation effects \cite{Galli2021}.% \\

Only a small number of works theoretically considered the effect of the surrounding spin-bath on a \textit{quantum register}, based on a microscopic evolution. For a two spin register in a P1-bath an expansion of the total evolution operator in terms of the Chebyshev-polynomials was used to model 15 bath-impurities \cite{vanderSar2012}. Furthermore, recently the entanglement to unwanted nuclear spins was taken in to account in \cite{Economou2023,Economou20232} for a multi-qubit NV-$^{13}C$-register. While the first approach is limited to a small number of bath-spins, the technique used in \cite{Economou2023,Economou20232} does not deal with spin-spin interactions within the environment. 

%%%%%%%%%%%%%%%%%%%%%%%%%%%%%%%%%%%%%%%%
In this work, we go substantially beyond these previous achievements in that we adapt the gCCE to simulate the dynamical performance of a complete spin register with up to four spins in diamond in the presence of interacting magnetic impurities. We further implement dynamical decoupling pulses offering the possibility to protect the evolution of the spin register from the static environment. 
This way, we also demonstrate the relevance of environmental spin-spin correlations and population transfer for multi-qubit registers and sequences of gate operations, and provide fidelity maps for four qubit quantum processes. 

More specifically, we analyze a quantum register consisting of one $\text{NV}^{-}$ (containing one electron spin and the nuclear spin of the nitrogen) and three $^{13}C$s embedded in a spin-bath consisting of nuclear- or P1-impurity spins, see Fig.~\ref{Fig.1}a.  
Further, we assume a purified diamond setup with $\eta_{^{13}C}\lesssim 0.05\%$
\footnote{In natural diamond the abundance  $\eta_{^{13}C} = 1.1\%$ yielding to a reasonable probability to find strongly coupled  $^{13}C$s ($\sim 100\,$MHz). Purified diamond (e.g. $\eta_{^{13}C} \approx 0.01\%$) leads to a sparse $^{13}C$ bath with better coherence properties for the NV spin.},
where nevertheless three $^{13}C$ are interacting individually with the $\text{NV}^{-}$ via dipole couplings with tens to hundreds of kilohertz as displayed in Fig.~\ref{Fig.1}a (see App.~\ref{App.A} for the dipolar-coupling strengths). 
Since we calculate the full evolution of the register starting from a chosen initialized quantum state and driven by pulse sequences as described in Fig.~\ref{Fig.1}b, i.e.\  far beyond the investigation of pure decoherence or relaxation phenomena, we obtain a comprehensive picture of realistic quantum operations.

We structure the article as follows:
In Sec.~\ref{Sec.2}, we start by introducing the full Hamiltonian of the register and the spin-bath. Then, we describe our theoretical modelling of the pulse structure in detail. Next, we a give short introduction into the gCCE. 
In Sec.~\ref{Sec.Results}, we first truncate the full register to a two qubit register and benchmark the gCCE with exact diagonalization for a quantum process in a spin-bath consisting of nine impurities. Then, we study the performance of different quantum processes on a fully connected four qubit register in presence of different spin-baths with dynamical decoupling pulses. We conclude in Sec.~\ref{Sec.Summary} and give an outlook how to overcome current limitations.

\section{Theory and Methods}\label{Sec.2}
\subsection{The Hamiltonian of the system}
We consider a spin register composed out of the $\text{NV}^{-}$ electron spin, the nuclear spin of the nitrogen atom and up to three nuclear $^{13}C$ spins.
The full Hamiltonian thus reads
\begin{align}
    H = H_R+ H_{\mathcal{B}} + H_{I},
\end{align}
where $H_{\mathcal{B}}$ is the Hamiltonian for the environment, $H_I$ the interaction Hamiltonian between register and environment and $H_R$ the one for the register containing three parts
\begin{align}
    H_R = H_{NV}+H_{N}+\sum_k H^{(k)}_{\text{C}}. 
\end{align}
For the purpose of this work, the bare $\text{NV}^{-}$ center can be readily described by the model Hamiltonian
\begin{align}
    H_{NV} = D_{gs}\, S_z^2 - g_e\,B_z S_z,
\end{align}
where $D_{gs}=2.88\,$GHz is the zero field splitting of the triplet state, $g_e=-2.806\,$MHz/G is the gyromagnetic ratio of the electron spin, and $B_z$ is an external magnetic field. We consider $B_z=148\,$G as in \cite{Suter2020} for reasons further explained below. 
We are interested in the effect of the environment on the quantum processes initiated by a microwave drive of the $\text{NV}^{-}$ spin.
As we will show later, these processes take several microseconds. In this regime the longitudinal relaxation to the spin environment is irrelevant.
This justifies a truncation of the $\text{NV}^{-}$ triplet state to a two - level system built out of the $m_s=0$ and the $m_s=-1$ state. In the following, we denote the matrices of the $\text{NV}^{-}$ spin 
\begin{align}
    \!\!S_z\!=\!\left(\!\!\begin{array}{cc}  0 &0\\ 0 & -1\end{array}\!\!\right),\, S_x\!=\!\frac{1}{\sqrt{2}}\left(\!\!\begin{array}{cc}  0 &1\\ 1 & 0\end{array}\!\!\right),\,S_y\!=\!\frac{1}{\sqrt{2}i}\left(\!\!\begin{array}{cc}  0 &1\\ -1 & 0\end{array}\!\!\right)\!.
\end{align}
Further, the Hamiltonian encoding the triplet nuclear spin state of the nitrogen ($I_N=1$) reads 
\begin{align}
    H_N= P_{gs}\;I_{N,z}^2 - g_n B_z I_{N,z} + \boldsymbol{S}\bar{N}\boldsymbol{I}_N
\end{align}
where $P_{gs}=-5.08\text{ MHz}$ is the zero field splitting, $g_n=0.308\,$kHz/G the gyromagnetic ratio and $\bar{N}$ the hyperfine coupling tensor between the nitrogen nuclear spin and the $\text{NV}^{-}$ electron spin. The bold quantities denote the full spin vectors.
Since we are only interested in the usage of the surrounding $^{13}C$ nuclear spins as quantum memory, we assume the nitrogen nuclear spin to remain without any dynamics through the whole process.  Because the energy splittings of nitrogen nuclear spin and the NV electron spin have different orders of magnitude the flip-flop interaction between the two spin species is suppressed. 
In this regime the nitrogen nuclear spin can be simply incorporated by using the Hamiltonian \cite{Suter2020}
\begin{align}
H_{NV}+H_N = D_{gs}\, S_z^2 +\left(- g_e\,B_z + \bar{N}_{zz} m_{N} \right) S_z,
\end{align}
where $m_{N} \in\{-1,0,+1\}$, e.g. we introduce the nuclear spin as renormalization of the Zeemann term for the electron spin. The NV-center can then be driven by microwave pulses within the subspace of a particular $m_N$, guaranteeing its frozen dynamics. 

The $^{13}C$s have nuclear spin $I_{C}=1/2$ and are described by the Hamiltonian
\begin{align}
    H^{(k)}_{C}=-g_C B_z I^{(k)}_{C,z} + \boldsymbol{S}\bar{M}^{(k)}\boldsymbol{I}^{(k)}_C + \boldsymbol{I}^{(k)}_{C} \bar{L}^{(k)} \boldsymbol{I}_{N},
\end{align}
where $g_C=1.071\,$kHz/G is the gyromagnetic ratio, $\bar{M}^{(k)}$ the hyperfine coupling between the $\text{NV}^{-}$ and the $k$-th $^{13}C$ and $\bar{L}^{(k)}$ is the hyperfine coupling between the $k$-th nuclear spin of the $^{13}C$ and the nitrogen nuclear spin. All dipole interaction tensors are further discussed in App.~\ref{App.A}.
We note that the dipole interactions between the nuclear spins are irrelevant on the timescales we discuss in this article. 
However, we take them into account in our simulation since they might become important for long sequences of quantum operations, or more dense registers,  we plan to simulate in the future. 

The register is assumed to be embedded in a finite size spin-bath consisting of either nuclear- spins or P1-impurities described by the Hamiltonian $H_{\mathcal{B}}$, where $\mathcal{B}\in\{\text{P1},^{13}C\}$ stands for the local Hamiltonian of the respective bath type (note that we denote a bath-spin operator irrespective of its type with $\boldsymbol{\mathcal{E}}$ in the following). Considering $^{13}C$-nuclear impurities the bath Hamiltonian is simply given by the sum of the Zeeman terms of the nuclear spins
\begin{align}
 H_{{^{13}}C}=-g_C B_z \sum_j\mathcal{E}^{(j)}_z.
\end{align}
The general Hamiltonian for the interaction of the environment with the register reads
\begin{align}
    \label{full-environment}
    H_{I} =& \sum_j \left(\boldsymbol{S}\bar{K}^{(j)} \boldsymbol{\mathcal{E}}^{(j)} + \boldsymbol{I}_N\bar{A}^{(j)} \boldsymbol{\mathcal{E}}^{(j)} \right. \\ & \left.  +\sum_k \boldsymbol{I}^{(k)}_C\bar{F}^{(k,j)} \boldsymbol{\mathcal{E}}^{(j)}   \right) + \sum_{ij} \boldsymbol{\mathcal{E}}^{(i)} \bar{G}^{(i,j)} \boldsymbol{\mathcal{E}}^{(j)}\nonumber,
\end{align}
where  $\bar{K}^{(j)} $ and $\bar{A}^{(j)} $ are the dipole interaction tensor between the bath spin (denoted by $\boldsymbol{\mathcal{E}}$) and the  $\text{NV}^{-}$ electron spin and its nuclear spin, respectively. 
$\bar{F}^{(k,j)}$ is the dipole coupling tensor between the $^{13}C$ nuclear spins of the register and the spin-bath and $\bar{G}^{(i,j)}$ the interaction within the spin environment.

\begin{figure}[t]
\includegraphics[scale=0.5]{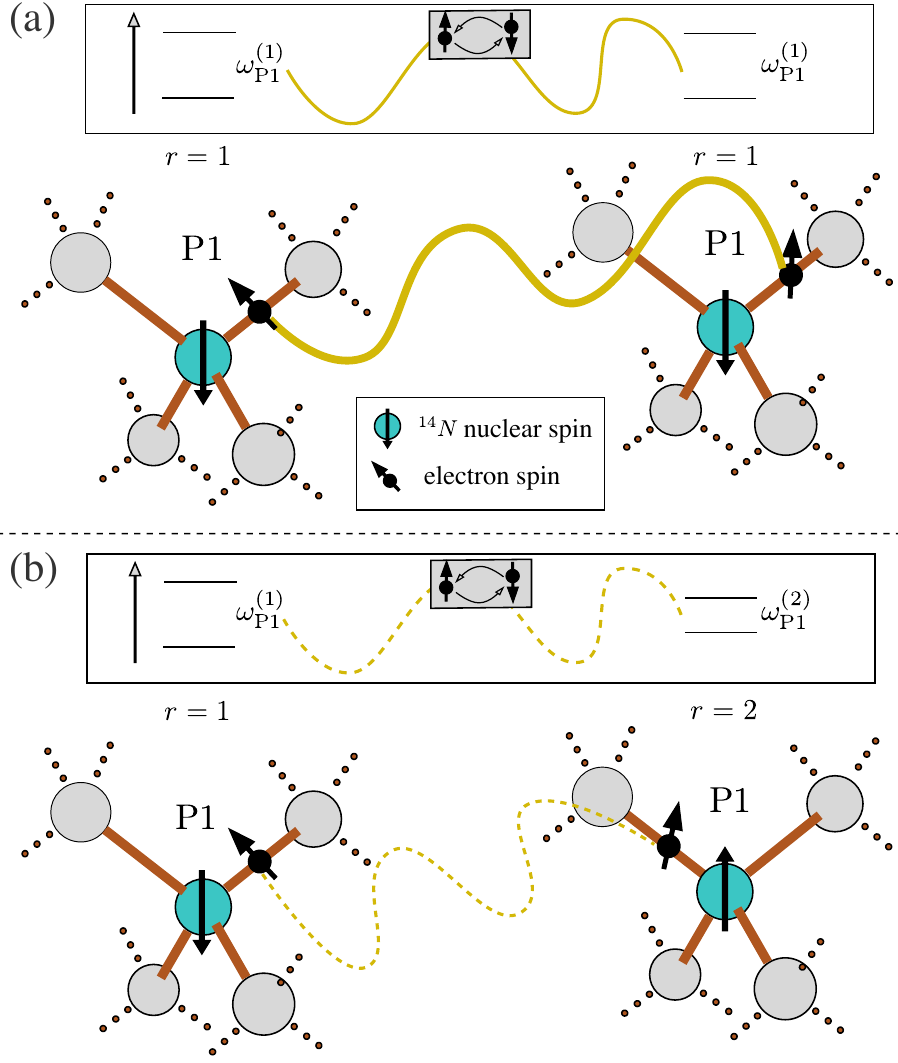}
%\scalebox{.8}{\input{Bild.tex}}
\caption{Two coupled P1-impurities. (a)  The delocalization axis and the state of the nitrogen nuclear spin coincide ($r=1$), hence the Larmor frequencies $\omega^{(r)}_{\text{P}1}$ of both centers coincide. This yields a strong contribution of the flip-flop part of the dipole interaction and to inherent dynamics within the spin-bath.
(b) The states of the P1-centers are different bringing the electron spins out of resonance and suppressing the flip-flop part of the dipolar interaction. }
\label{Fig.1bb}
\end{figure}

For the local Hamiltonian of a P1-impurity the electron spin and the nuclear spin of the nitrogen has to be taken into account. The local Hamiltonian for impurity $j$ reads
\begin{align}
H_{\text{P1}}^{(j)} &=   - g_e B_z \mathcal{E}^{(j)}_z + g_n B_z \mathcal{I}^{(j)}_z + A_T \mathcal{E}_z^{(j)} \mathcal{I}_z^{(j)} \nonumber \\ &+ A_L \left(\mathcal{E}_x^{(j)} \mathcal{I}_x^{(j)} + \mathcal{E}_y^{(j)} \mathcal{I}_y^{(j)} \right) - P_{N} (\mathcal{I}_z^{(j)})^2,\label{Eq.P1-Ham}
\end{align}
where $A_T$ is the transversal and $A_L$ the longitudinal coupling between the electron and the nitrogen nuclear spin ($\boldsymbol{\mathcal{I}}$) and $P_N=-4 $ MHz the zero field splitting of the nuclear spin. For large enough magnetic field the flip-flop interaction between the two spin species is suppressed. Further, because only the dynamics of the electron spin play a role for the register, we can write the Hamiltonian as
\begin{align}
H_{\text{P1}}^{(j)} &\approx   - g_e B_z \mathcal{E}^{(j)}_z + A_T \mathcal{E}_z^{(j)} \mathcal{I}_z^{(j)}.
\end{align}
The hyperfine coupling strength $A_T$ depends on the specific delocalization axis of the P1-center in the diamond lattice.
Particularly, we use $A_T=114$ MHz for the $[111]$ axis and $A_T = 86$ MHz for the $[\bar{1},1,1 ]$, $[1,\bar{1},1 ]$, $[1,1,\bar{1} ]$ axes \cite{Witzel2012}. We model the effective P1 Hamiltonian by treating the effect of its nuclear spin as a local random magnetic field contribution
\begin{align}
H_{\text{P1}}^{(j,r)} &\approx \left( \underbrace{- g_e B_z  + A_T^{(r)} m^{(r)}_{N,P1}}_{\omega^{(r)}_{P1}}  \right) \mathcal{E}_z^{(j)} .
\label{Eq.disorder}
\end{align}
where $r$ is the index for the delocalization axis and the state of the nuclear spin given by $m^{(r)}_{N,P1} \in \{-1,0,1\} $. Since, we assume a completely mixed thermal state in the following, we sample over the twelve different configurations for $r$.
The local disorder in the magnetic field brings the P1 impurities out of resonance  yielding a suppression of P1-P1 interactions within the spin-bath (see also Fig.~\ref{Fig.1bb}). 
As described above, we work in the regime where $B_z=148$ G, which is not completely suppressing the inherent P1 dynamics governed by Eq.~(\ref{Eq.P1-Ham}). However, the amplitude of oscillations of $\langle \mathcal{E}_z\rangle $ due to the flip-flop interaction with the onsite nuclear spin remains small with $\langle \mathcal{E}_z(t)\rangle \in [0.45,0.5]$ and we assume that the effects of this oscillation on the coherence properties of the register is of higher order. In this way our results for the P1-spin-bath can be seen as an upper bound for the fidelity of the simulated quantum processes.

%%%%%%%%%%%%%%%%
%%%%%%%%%%%%%%%%

%
\subsection{Quantum processes on the spin register}\label{Sec.optimization}
The operations on the quantum register are initiated solely by driving the electron spin of the NV center. We follow  protocols suggested by several experiments in this direction \cite{Suter2020,Suter2020-2}. Particularly, we consider pulse sequences of the form shown in Fig.~\ref{Fig.1}b. 
The full unitary evolution operator of the register is encoded in the parameters $\{t_i,\theta_i,\phi_i\}$. Here, the angles $\theta_i$ and $\phi_i$ describe a particular rotation on the Bloch sphere of the NV spin and $t_i$ are times for which we let the system interact. 
To find the parameters of the particular process, we encode the sequence of pulses into subsequent unitary operations
\begin{align}
    \mathcal{S}(\boldsymbol{t},\boldsymbol\theta,\boldsymbol\phi)\!=\!... U_{\theta_4,\phi_4}U_{t_3}U_{\theta_3,\phi_3}U_{t_2}U_{\theta_2,\phi_2}U_{t_1}U_{\theta_1,\phi_1}U_{t_0},
\end{align}
where $U_{t_i}$ denotes the unitary evolution with the  Hamiltonian $H$ (generally containing the spin-bath) and 
\begin{align}
   U_{\theta_j,\phi_j} = e^{-i\frac{\theta_j}{2}\left(\sigma_x\otimes \mathbb{1}_v\cos(\phi_j)+\sigma_y\otimes \mathbb{1}_v\sin(\phi_j)\right)}
\end{align}
is encoding the pulse. $\sigma_x$ and $\sigma_y$ are Pauli matrices acting on the NV-electron spin and $\mathbb{1}_v$ is the identity matrix for the rest of the system. We optimize the infidelity $\mathcal{I}_p=1-\mathcal{F}_p$, where $\mathcal{F}_p$ is the fidelity of the state after the gate sequence \textit{without} the dissipative couplings
\begin{align}
    \rho_p = \mathcal{S}(\boldsymbol{t},\boldsymbol\theta,\boldsymbol\phi)\, \rho(0) \,  \mathcal{S}^\dagger(\boldsymbol{t},\boldsymbol\theta,\boldsymbol\phi)
\end{align}
with respect to the target state $\rho_T$ defined as
\begin{align}\label{Eq.firstfid}
    \mathcal{F}_p=\frac{\text{tr}\left(\rho_T^\dagger \rho_p\right)}{\text{tr}\left(\rho_T^\dagger \rho_T\right)}.
\end{align}
Note that $\rho_p$ and $\rho_T$ are density matrices of the full spin register. We assume the pulse unitaries $U_{\theta_i,\phi_i}$ to be instantaneous and perfect.
Further, we adjust the magnetic field in a way, that the bare Larmor frequencies $|\omega_C| = |g_C B_z|$ of the register-$^{13}C$s are of the order of the coupling to the NV electron spin. Then, when the NV is brought into the $m_s = -1$ state, the quantization axis of the $^{13}C$ becomes almost perpendicular to the z-direction. This can be easily seen by projecting the Hamiltonian of the register into the $m_s=-1$ state. We obtain an effective Hamiltonian for the evolution of one nuclear spin (here $k=1$) by neglecting irrelevant flip-flop terms with the NV
\begin{align}
H^{(m_s=-1)}_{R,1}&\approx- (g_{C} B_{z}+\bar{{M}}_{zz}^{(1)}) {I}^{(1)}_{C,z}\!-\bar{{M}}_{zx}^{(1)} {I}^{(1)}_{C,x} \nonumber
 \\
&\equiv-\epsilon_0 {I}^{(1)}_{C,z} -\Delta {I}^{(1)}_{C,x}.
\end{align}
In the limit $\epsilon_0\sim \Delta$ (or even $\epsilon_0\ll\Delta$ because $\bar{M}^{(1)}_{zz} = -153$ kHz  and $g_C B_z= 158 $ kHz for $k=1$ see App.\ref{App.A}) the quantization axis of the nuclear spin is strongly tilted from the z-direction. 
In this way, one can address the nuclear spins 
with only a few pulses. We test this scheme - originally proposed in \cite{Suter2020} for single spin addressing - for its ability to be generalized to coherently address several nuclear spins at the same time via their common dipole interaction with the NV-center. This scheme is similar to the one implemented in \cite{Taminiau2014} but in the above explained magnetic field regime, reducing the number of pulses and therefore shortening the processes significantly. 
Since the main focus of this paper is on the simulation of a full quantum register within a spin-environment, we do not claim that the pulse sequences introduced are the most sophisticated ones. However, we find a regime where such operations can be very efficient. They particularly  require isotopically purified diamond samples as we will also explain in the following. 
Furthermore, we propose that different control schemes can be efficiently tested against spin-environments using the technique we introduce in the next section. 

In Sec.~\ref{Sec.DD}, we additionally implement dynamical decoupling (DD) sequences to protect the quantum process from the environment. There, we interleave each unitary evolution $U_{t_i}$ with one \textit{hahn-echo} pulse
\begin{align}
    \mathcal{S}_{\text{DD}}(\boldsymbol{t},\boldsymbol\theta,\boldsymbol\phi)= \!... U_{\theta_2,\phi_2}U_{t_1/2}U_{\pi,0}U_{t_1/2}U_{\theta_1,\phi_1}U_{t_0}.
\end{align}
This structure is further explained in Fig.~\ref{Fig.pulsesdd}. 
In both cases (with and without DD), we use the python build in function \textit{scipy.basinhopping} for the optimization of the free parameters. 

\subsection{Generalized Cluster Correlation Expansion}
To efficiently simulate the effect of the environment on the spin register, we use the generalized Cluster Correlation Expansion (gCCE) taking the full system Hamiltonian for the unitary evolution into account. 
 Within this methodology, higher order correlations within the environment are considered by subsequently calculating higher order terms in the expansion.  
We start with the well-established zeroth order (gCCE0) corresponding to the mean field approximation and yielding the simplified bath Hamiltonian
\begin{align}
    H^{(\text{E0})}_{I} =& \sum_j \left( S_z\bar{K}_{zz}^{(j)} \tilde{\mathcal{E}}_z^{(j)} + I_{N,z}\bar{A}_{zz}^{(j)} \tilde{\mathcal{E}}_z^{(j)} \right.\nonumber \\ & \left.  +\sum_k I^{(k)}_{C,z}\bar{F}_{zz}^{(k,j)} \tilde{\mathcal{E}}_z^{(j)}   \right),
\end{align}
where the bath spins $\tilde{\mathcal{E}}_z^{(j)}=\pm \frac{1}{2}$ are no longer operators, but chosen by classical probability (static disorder) and $\text{E0}$ stands for gCCE0. 
From now on, a particular collective bath-state (e.g. $\{\tilde{\mathcal{E}}_z^{(1)},\tilde{\mathcal{E}}_z^{(2)},\tilde{\mathcal{E}}_z^{(3)},..,\tilde{\mathcal{E}}_z^{(N)}\}=\{\frac{1}{2},-\frac{1}{2},-\frac{1}{2},..,\frac{1}{2}\}$) is encoded by the superscript $n=1,\ldots 2^N$ counting all possible reservoir configurations.
Considering temperatures by far exceeding the Zeeman splitting of the P1-impurities ($k_B T /\hbar\gg 400$ MHz), we assume the states of the bath spins to be random, particularly, in this work, we will use a completely mixed thermal bath-state $\rho_B= \mathbb{ 1}_B/2^N$ \cite{LiuReview}. 
The so called gCCE0 contribution to the density matrix reads therefore
\begin{align}
    \rho^{(n)}_{\text{E0}}(t_e) = \mathcal{S}_n(\boldsymbol{t},\boldsymbol\theta,\boldsymbol\phi)\rho(0)\mathcal{S}_n^\dagger(\boldsymbol{t},\boldsymbol\theta,\boldsymbol\phi).
\end{align}
The unitary operator of the systems evolution within the mean field bath is 
\begin{align}
    U_{t,n}=\text{exp}\left(-i H^{(E0,n)} t\right),
\end{align}
where $n$ labels a specific bath configuration and $H^{(E0,n)}=H_{R}+H_{I}^{(E0,n)}$. For simulating a completely mixed density matrix, we need to average over $M$ different random bath-states yielding
\begin{align}
    \rho_{\text{E0}}(t)=\frac{1}{M}\sum_{n=1}^M     \rho^{(n)}_{\text{E0}}(t).
\end{align}
The next order (gCCE1) is then calculated by additionally taking the full interaction between one specific bath spin $l$ and the system into account, see Eq.~(\ref{full-environment}). For this, we start with the initial state
\begin{align}
    \tilde{\rho}^{(n,l)}_{\text{E1}} (0)= \rho(0) \otimes \rho^{(n)}_{l}(0),
\end{align}
where $n$ denotes the bath configuration and $l$ the spin we add to the density matrix. 
The time evolved density matrix of the system for the specific spin $l$ reads
\begin{align}
        \rho^{(n,l)}_{\text{E1}}(t_e) = \text{tr}_{B,l}\left(\mathcal{S}_{n,l}\,    \tilde{\rho}^{(n,l)}_{\text{E1}}(0)\, \mathcal{S}_{n,l}^\dagger \right).
\end{align}
Note that the pulse sequence operators $\mathcal{S}_{n,l}$ now also contain the full interaction of the spin $l$ in the environment and its local Hamiltonian (encoded in $H_\mathcal{B}^{(E1,n,l)}+H_I^{(E1,n,l)}$) with the register. 
With the definition of these quantities, a density matrix element of the full gCCE1 contribution is defined as
\begin{align}
   \langle a | \rho_{\text{E1}}^{(n)} (t) | b\rangle = \langle a | \rho^{(n)}_{\text{E0}}(t)| b\rangle\,\prod_l  \frac{\langle a|\rho^{(n,l)}_{\text{E1}}(t)|b\rangle}{\langle a | \rho^{(n)}_{\text{E0}}(t)| b\rangle}, \label{Eq.dividing}
\end{align}
where each numerator in the product contains the full interaction with a different environmental spin. Since $\rho^{(n,l)}_{\text{E1}}(t)$ contains also the mean field contribution for all other spins apart from $l$, we have to divide by it for each term to only multiply corrections to the mean field approximation. The final gCCE1 density matrix is then calculated via
\begin{align}
    \rho_{\text{E1}}(t)=\frac{1}{M}\sum_{n=1}^M     \rho^{(n)}_{\text{E1}}(t).
\end{align}
With this framework also higher order correlations within the spin-bath can be taken into account, as we show in App.~\ref{App.B}.
In the following, we use this method to calculate the density matrix of the full spin register influenced by the environment. To analyze the performance in presence of dissipation, we define a second fidelity
\begin{align}
    \langle\mathcal{F}_f\rangle=\frac{\text{tr}\left(\rho_p^\dagger \rho(t_e)\right)}{\text{tr}\left(\rho_p^\dagger \rho_p\right)},
\end{align}
encoding the deviation from unitary evolution $\rho_p$ due to the coupling with the bath \cite{Chen_2020}. Note that $\rho(t_e)$ denotes the exact solution for the reduced density matrix of the register at the end of the pulse sequence in presence of the environment and will be approximated by different gCCE orders in the next sections. 
For the four spin register depicted in Fig.~\ref{Fig.1}a the numerical calculation of the full density matrix using the gCCE becomes  challenging. The main problem are elements that are close to zero yielding diverging entries stemming from dividing by zero in Eq.~(\ref{Eq.dividing}).
However, for the calculation of the fidelity, not the full density matrix is needed, in these situations we perform the gCCE directly on $ \mathcal{F}_f$. For more information,  also to the accuracy of the gCCE, we refer to App.~\ref{App.B} .

Although the calculation of the full fidelity $\langle \mathcal{F}\rangle=\mathcal{F}_p \langle\mathcal{F}_f\rangle$ is straight forward, we will mainly focus on the distinct effect of the spin-bath and therefore on $\langle \mathcal{F}_f \rangle$ in the results section.
We find that for short times the gCCE0 contribution, i.e. treating the spin-bath in the mean field approximation, is sufficient (e.g. $\rho(t)\approx \rho_{\text{E0}}(t) $). However, we also analyze under which circumstances higher order terms become necessary in the next section (e.g. $\rho(t)\approx \rho_{\text{E2}}(t) $).  
\begin{figure}[b]
\includegraphics[scale=0.41]{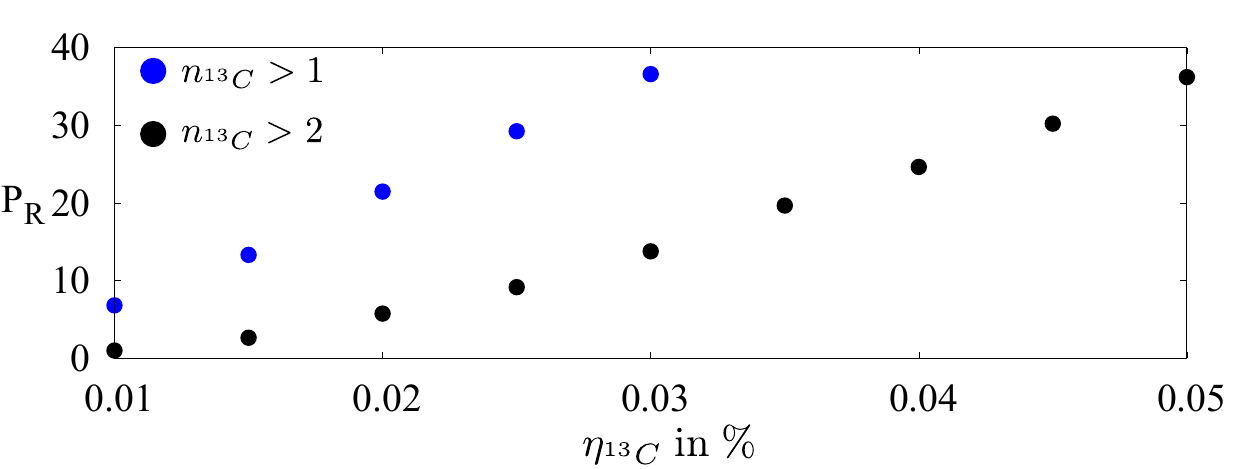}
%\scalebox{.8}{\input{Bild.tex}}
\caption{ Probability to find a number of at least $n_{^{13}C}$ nuclear spins within a range of $r=1.8$ nm around the NV-center for different abundances $\eta_{^{13}C}$. The blue dots correspond to the probability to find at least two, the black dots to find at least three spins. The points are calculated by randomly producing 20000 spin-bath configurations within a sphere of radius $r_{max}=10$ nm. }
\label{Fig.2n}
\end{figure}

\begin{figure*}[t]
\includegraphics[scale=0.63]{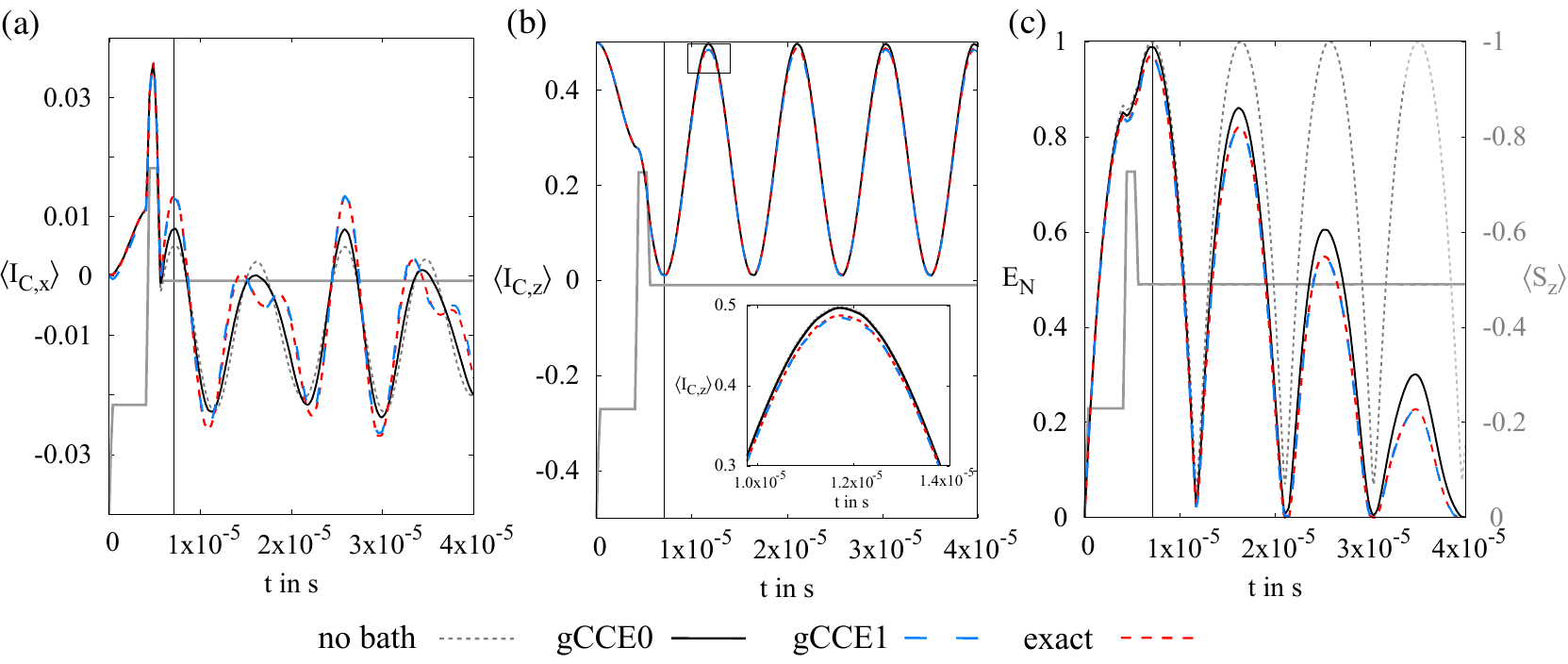}
%\scalebox{.8}{\input{Bild.tex}}
\caption{(a) Influence of the pulse sequence on the $I_{C,x}$ component of the $^{13}C$ nuclear spin. The vertical line shows the end of the gate operation $t_e$. (b) $z$-component of the $^{13}C$ spin. The inset shows a zoom into the small square region of the plot visualizing the change in the amplitude of the oscillation of $I_{C,z}$ for gCCE1 and gCCE0. The decreasing amplitude signals energy flow to residual nuclear spins during the pulse sequence. (c) Logarithmic negativity $E_N$ as a measure of the entanglement in the system.  The deviation of the dissipative evolution from the unitary is a measure for the fidelity of the process. The deviation grows for longer times. Each gCCE evolution was averaged over all possible bath-states. The grey line in each plot shows the $S_z$ component of the NV electron spin under the influence of the pulse sequence. The ticks for   $S_z$ are displayed on the right axis of (c).  }
\label{Fig.2}
\end{figure*}
\section{spin-bath properties}

\subsubsection{The $^{13}C$-nuclear-spin-bath}
In this paper, we assume a purified diamond sample with $\eta_{^{13}C}\lesssim0.05\%$ abundance of $^{13}C$ isotopes with up to three $^{13}C$ nuclear spins being close to the NV-center.  We further assume, that we address every $^{13}C$ nuclear spin within a spherical shell of $r_{min}=2$ nm. Nuclear spins that are further away become hard to initialize and control due to the small- and sub-kiloherz coupling to the NV-center. Hence, we consider that the first unwanted nuclear spin is at least $r=2$ nm away from the electron spin. 
We show the probability of finding a register with up to three $^{13}C$ nuclear spins in the vicinity of the electron spin (see also Fig.~\ref{Fig.1}a) depending on the abundance $\eta_{^{13}C}$ in Fig.~\ref{Fig.2n}. 
In such a sparse nuclear spin-bath the $^{13}C$s interact only weakly with each other and we will find that the dynamics stemming from intra-bath interactions are irrelevant for the simulated timescales. 
\subsubsection{The P1-spin-bath}
As mentioned in Sec.~\ref{Sec.2} the modelling of a P1-impurity-bath is more challenging since the local nitrogen nuclear spin of each P1-center has to be taken into account.
We model its effect as a local disorder field which is sampled over. The gCCE is perfectly suited for this purpose since we have to average over different bath states $M$ in the first place. 

In principle, P1-impurities at a distance $r<30\,$nm from the NV yield interaction strengths comparable to the $^{13}C$ nuclear spins. Without dynamical decoupling pulses, such a situation would destroy the register in the fist place. Accordingly, in the following subsection we first assume a situation, where residual bath-spins reside outside a central core of $r_{\text{min}}=30$ nm (see Fig.~\ref{Fig.1}a).
In Sec.~\ref{Sec.DD}, we implement dynamical decoupling pulses and weaken this assumption by decreasing the core size to $r_{\text{min}}=10$ nm.
In presence of decoupling pulses we study dense P1-spin-baths with $\eta_p>200$ ppb. 

We want to stress that purified diamond devices with $\eta_{P1}\approx 5$ ppb are available and have been used to perform quantum algorithm experiments. 
However, in such low impurity samples the necessary nitrogen implantation also creates impurities (P1, vacancies, divacancies) also leading to dynamics within the spin-bath. The discussed framework is suitable to also address such questions.
Furthermore, we point out that we are considering the local impurity density around the NV-center and a spin-bath with $\eta_{\text{P1}}=200$ ppb has on average six P1-centers within a radius of $35$ nm, comparable to a spin-bath that might be created by implantation techniques aiming for the creation of several NV-$^{13}C$-registers close to each other \cite{Shinobu2019}.
Because the close pairs of P1-spins have the most detrimental effect on the processes, especially the next subsection gives insight in purified diamond samples, where the P1-bath (here nine impurities) is created by implantation. 

\section{Results}\label{Sec.Results}

\subsection{Benchmark: Entanglement gate between the NV and one $^{13}C$ in the presence of a small spin-bath} \label{Sec.twospins}
Before we discuss the full system, we benchmark the gCCE via exact diagonalization of a two qubit spin register within a spin-bath consisting of nine impurities. Thereby, we assume only one $^{13}C$ to be close to the $\text{NV}^{-}$ comparable to the experimental system of Ref.~\cite{Suter2020} and described by the effective register Hamiltonian via
\begin{align}
\!\!H_R\!=\!D_{gs} S_{z}^{2}\!-\!\left(g_{e} B_{z}\!-\!N_{zz}\right) S_{z}\!+\!\boldsymbol{S} \bar{{M}}^{(1)} \boldsymbol{I}^{(1)}_{C}\!-\!g_{C} B_{z} I^{(1)}_{C,z},
\end{align}
where we assume that the nuclear nitrogen spin remains in the state $m_N = -1$ and $N_{zz} = - 1.76$MHz is its diagonal dipole coupling to the NV electron spin \cite{Korener2021}. 

In the following, we simulate an entanglement gate between the NV and $^{13}C^{(1)}$ in the presence of a small spin-bath (see App.~\ref{App.C} for the parameters of the pulse sequence). 
We start with a sparse bath of $^{13}C$ nuclear spins and show the evolution of the sequence in Fig.~\ref{Fig.2}.
In (a) and (b), we display the evolution of the $x$- and $z$-component of the nuclear spin in the register, due to the pulse sequence on the NV electron spin (the grey line in each plot shows the $S_z$ component of the NV under influence of the pulse sequence).
In (c), we show the logarithmic negativity ($E_N$) as a measure for the entanglement in the system. This quantity is suitable to quantify the  bipartite entanglement also for mixed density matrices and is defined as
\begin{align}
    E_{N}[{\rho}]=\log _{2}\left(\left\|{\rho}^{T_{A}}\right\|_{1}\right),
\end{align}
where $\left\|...\right\|_1$ is the trace norm and ${\rho}^{T_{A}}$ the partially transposed density matrix with respect to one subsystem (here the $^{13}C$ nuclear spin)\cite{Werner2002,Plenio2005}.
%

%%%%%%%%%%%%%%%%%%%%%%%%%%%%%%%%%
\begin{figure*}[t]
\includegraphics[scale=0.64]{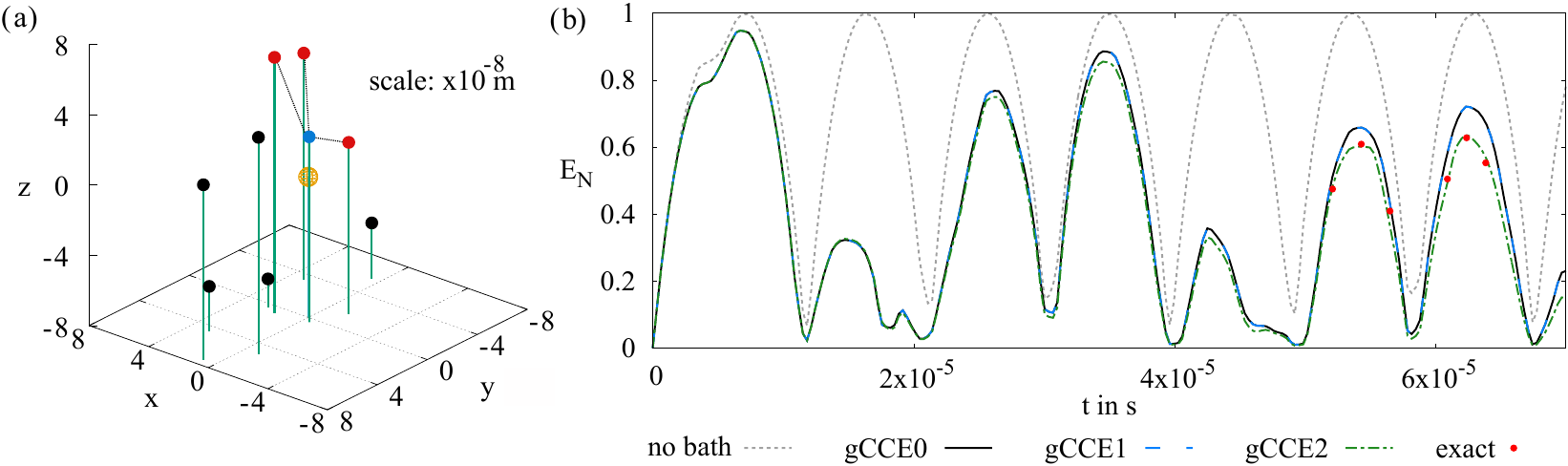}\\
%\includegraphics[scale=0.35]{logneg-long-CCE0.eps}\\
%\includegraphics[scale=0.47]{logneg-CCE2.pdf}
%\scalebox{.8}{\input{Bild.tex}}
\caption{(a) Spin-bath configuration used for plot (b). The orange sphere represents the register while the remaining dots represent the bath spin. The green lines are guides for the eye.  (b) Long time behavior of the logarithmic negativity for the spin-bath of (a). The revivals of the dissipative evolution are correctly captured by the gCCE2 approximation. Here, only the correlation between the connected pairs of (a) are taken into account (blue and red dots). The short time behavior of the evolution is sufficiently described by gCCE0. The gCCE1 contribution coincides with the gCCE0, because the NV center does not induce any dynamics into the spin-bath. Each data point was averaged over 1050 different bath-states (M=1050).   }
\label{Fig.3}
\end{figure*}
%%%%%%%%%%%%%%%%%%%%%%%%%%%%%%%%%%%
%
%
We find that at the end of the pulse sequence ($t_e \approx 7.1\, \mu s$ vertical line) $E_N \approx 1$ signaling the success of the process in the case without an environment (grey dashed line).
The unitary evolution shows coherent revivals of the gate-induced entanglement in the system for $t>t_e$.

Considering the bath, we distribute nine impurity spins equally  within a spherical shell with $r_{min} = 2.0$ nm and $r_{max} = 4.2\,$nm. This corresponds on average to a volume density of $\eta_{^{13}C} \approx 0.02\, \%$. 
The red dashed line corresponds to the result obtained via exact diagonalization. 
In Fig.~\ref{Fig.2}a, we obtain that the evolution of the $x$-component of the nuclear spin in the register is strongly altered by the presence of the environment. While the gCCE0 only affects the amplitude of the oscillations the gCCE1 resolves also additional contributing frequencies within the evolution for $t>t_e$. Furthermore, the magnitude of oscillation of the $z$-component of the nuclear spin after the pulse sequence is reduced with respect to non dissipative system, indicating population transfer to unwanted nuclear bath spins (see Fig.~\ref{Fig.2}b). 
Together with the classical meanfield contribution this leads to a reduced fidelity of the quantum process as can be seen from the deviation of the gCCE0 and gCCE1 contribution to the non dissipative evolution.
Furthermore, we observe the decay of the entanglement between the NV-center and the $^{13}C$ spin after the pulse sequence $t>t_e$ (see Fig.~\ref{Fig.2}c).  
We find that the gCCE1 contribution describes the result obtained by exact diagonalization sufficiently well. Hence, spin-spin interactions within the nuclear spin-environment are irrelevant on the simulated timescales.  
Regarding the P1-center spin-bath, we first look at very sparse abundances $\eta_{\text{P1}}$. 
To benchmark the above introduced gCCE, also for this kind of bath, we show the evolution for longer time scales to observe at which point higher order terms become important.
Especially for a spin-bath consisting of P1-impurities correlations can build up fast with respect to the timescales of the process on the register, because the electron spins are strongly interacting.
In Fig. \ref{Fig.3}b, we show the long time evolution of the logarithmic negativity in a P1-spin-bath. In Fig.~\ref{Fig.3}a we show an example for a random distribution of nine P1-impurities within a spherical shell with $r_{min}=30\,$nm and $r_{max}=80\,$nm corresponding on average to an abundance $\eta_{\text{P1}} \sim 26\,$ppb.
First of all, we see that the gCCE1 contribution coincides with the gCCE0 meaning that the pulse sequence does not induce any dynamics into the P1-spin-bath. This can be easily explained by their fact that the initialized quantum process is designed to affect nuclear spin having a much lower Larmor frequency. 
At a certain point in time the gCCE2 contribution - also taking correlations within the spin-bath into account - describes the evolution better then gCCE0 and gCCE1 and concides with the exact diagonalization \footnote{Note that we implemented the local Hamiltonian (\ref{Eq.disorder}) for the P1-impurities also for the exact calculation, which we also need to sample over 300 different local disorders.}.
Particularly, we obtain that the revivals of the entanglement, are only correctly simulated when we account for the growth of correlations in the environment. 
For this specific case it is enough to include the correlations between the closest pairs of bath spins (highlighted in Fig. 3a by the dotted lines), since correlation between spins that are further away builds up on longer time scales.
We performed this selection by only taking the gCCE2 contribution of spins that are closer than $55\,$nm and have a second spin within $55\,$nm in their surrounding. 
Reversing the discussion - for a small number of spins in the environment - the gCCE can also be used as a tool to analyze a given spin-bath and to find which specific correlations become important for the register on different timescales. 
\begin{figure}[b]
\includegraphics[scale=0.42]{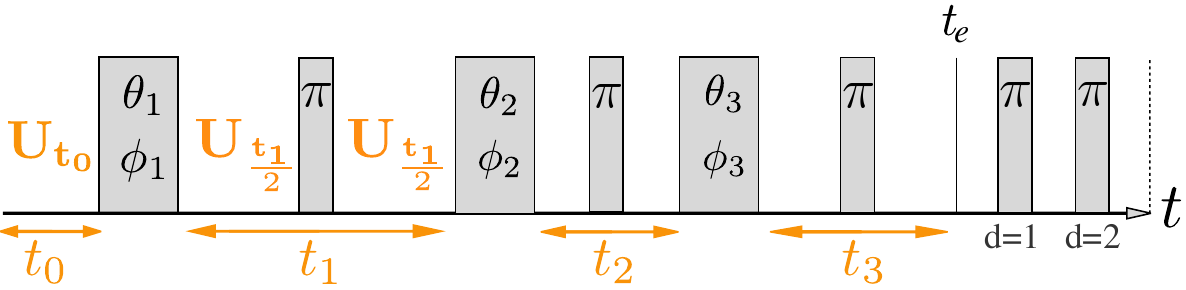}
%\scalebox{.8}{\input{Bild.tex}}
\caption{Pulse sequence including dynamical decoupling pulses. We insert one $\pi$-pulse in the middle of every free evolution to decoupling the gate from the surrounding environment. For times $t>t_e$, we protect the established information with additional decoupling pulses (numbered by $k$). }
\label{Fig.pulsesdd}
\end{figure}

\begin{figure*}[t]
\includegraphics[scale=0.475]{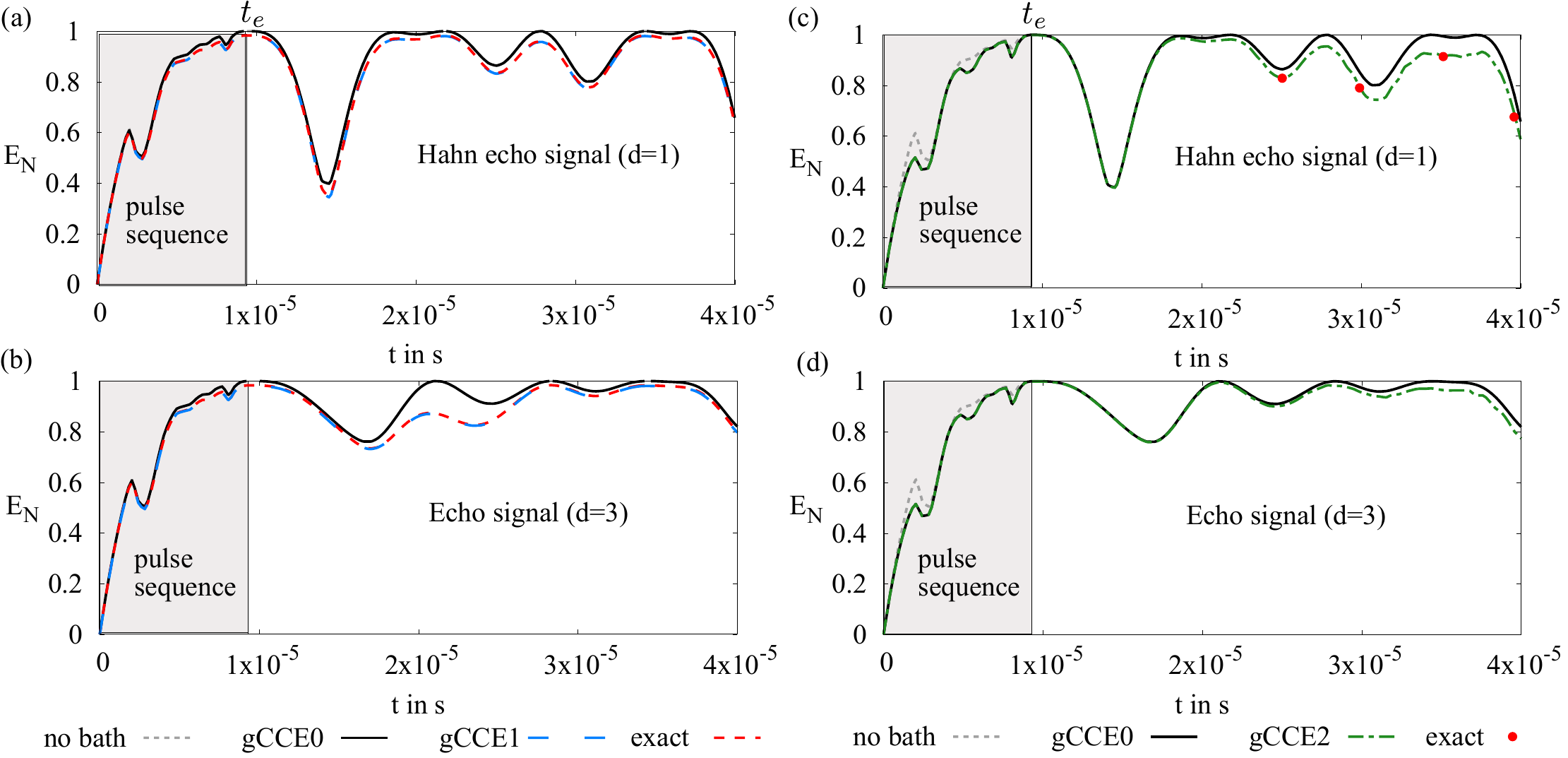}
%\scalebox{.8}{\input{Bild.tex}}
\caption{Protected entanglement process between the NV electron spin and the $^{13}C$ nuclear spin. The grey area in (a)-(d) corresponds to the timescale of the entangling pulse sequence ($t \leq t_e$). After the sequence, different numbers of decoupling pulses $d$ protect the quantum information ((a) $d = 1$, (b) $d = 3$). (a) and (b) are calculated in a nuclear spin-bath. The red dashed line corresponds the exact result of the pulse sequence including dynamical decoupling, within a spin-bath of nine impurities ($\eta_{^{13}C} \approx 0.05 \%$). The black line displays the unitary evolution which coincides with the gCCE0 and the gCCE1 in this case. 
(c) and (d): Protected entanglement gate in the presence of a P1-spin-bath for $d=1$ in (a) and $d=3$ in (b). The gray dashed line shows the unitary evolution during the pulse sequence. The influence of the mean field contribution is canceled out for $t\geq t_e$. The green dashed-dotted line shows the gCCE2, which is in good agreement with the exact results (see (c)). Hence, the onset of pair-correlations within in spin-bath leads to a deviation from the unitary evolution. The results for times $t>t_e$ are echo signals after $d$ decoupling pulses for every $t$. For larger numbers of decoupling pulses the dissipative evolution remains comparable the the unitary evolution for longer times.   } 
\label{Fig.50}
\end{figure*}

\subsubsection{Protection of quantum processes via dynamical decoupling}\label{Sec.DD}
For larger abundances and on long timescales dynamical decoupling (DD) sequences are necessary to obtain fidelities $\langle\mathcal{F}_f\rangle>95\,\%$ for the quantum processes described above. 
In the NV-based spin register, we show in Fig.~\ref{Fig.1}a, the electron spin is the limiting factor for the fidelity, since its coupling to the unwanted impurities is stronger than the coupling of the $^{13}C$s to the environment.
Hence, a minimal strategy to protect the process is to use dynamical decoupling pulses for the electron spin only. 
To find protected pulse sequences, we again use the optimization procedure as explained in Sec.~\ref{Sec.optimization}. However, here, we force $\pi$-pulses to be in the middle of each unitary evolution, as explained in Fig.~\ref{Fig.pulsesdd}. The parameters of the optimized pulse sequence are again summarized in the App.~\ref{App.C}. 
Independent of the time in between the decoupling pulses, such a DD-protocol protects the sequence entirely from the static disorder. 
Generally, apart from decoupling from the environment, for a suitable choice of the times $t_i$ the DD-pulses also partly decouple the electron spin of the NV - center from the other nuclear system spins in the register (see next section). This has the effect that individual spins (or spin pairs) can be more efficiently addressed \cite{Plenio2015,Economou2023}, yielding more precise processes and therefore larger fidelities $\mathcal{F}_p$, especially in the four spin register. 
In the present case, considering only one nuclear spin being part of the register, we find for the non dissipative case $1-\mathcal{F}_p < 10^{-5}$.

We show the result for the logarithmic negativity of the two spin register in Fig.~\ref{Fig.50}~a-b. There, we again compare the gCCE with the exact solution for the same spin-bath as in Fig.~\ref{Fig.2}. 
We find that the process, which takes $t_e \approx 9.3\,\mu$s, is protected from the static noise of the environment. This can be seen by the fact that the gray dashed line and the black solid line coincide. 
However, the additional DD-pulses during the gate sequence do not necessarily decouple from the surrounding $^{13}C$ nuclear spins. The population transfer is still present and can be on the same order or even worse compared to the process without dynamical decoupling. 
In the presented case the protected process performs slightly better $\langle \mathcal{F}_f \rangle \approx 0.98 $ then the unprotected one $\langle \mathcal{F}_f \rangle \approx 0.97 $ (refering to Fig.~\ref{Fig.2}c). 
This transfer reveals the caveat of the used pulse sequence. Since it is designed to strongly address the nuclear spin with only a few pulses, its \textit{impact radius} is difficult to control. This problem is not so apparent in more commonly used pulse sequences  \cite{Plenio2015,Taminiau2014,Economou2023}. They work in different magnetic field regimes and take usually more time, with up to hundreds of pulses where one has to deal also more carefully with control errors.
The evolution for $t>t_e$ is protected from additional decoherence effects by using Hahn echo pulses. Although Fig.~\ref{Fig.50}c - having three decoupling pulses after the pulse sequence - reveals a larger error at $t\approx 20\,\mu$s indicating a resonance with the spin-bath and therefore a larger population transfer. 

In Fig.~\ref{Fig.50}~c-d, we show the same protocol within a P1-spin-bath with $\eta_{\text{P1}}\approx 400$ ppb. Here, we find that although the dissipative evolution at times $t<t_e$ deviates from the non dissipative one, the process is protected at $t=t_e$.  After the pulse sequence, we aim to protect the evolution of the entanglement via different numbers of decoupling pulses $d$ (see also Fig.~\ref{Fig.pulsesdd}). For longer times, the correlations within the spin-bath yield a small deviation from the unitary evolution. We find that gCCE2 approximation describes the exact solutions well on these timescales, suggesting that higher order correlations are not important. We further find, that inserting more DD pulses protects the evolution for $t>t_e$ better from the interactions within the P1-environment (see (d)). 

\begin{figure*}[t]
\includegraphics[scale=0.48]{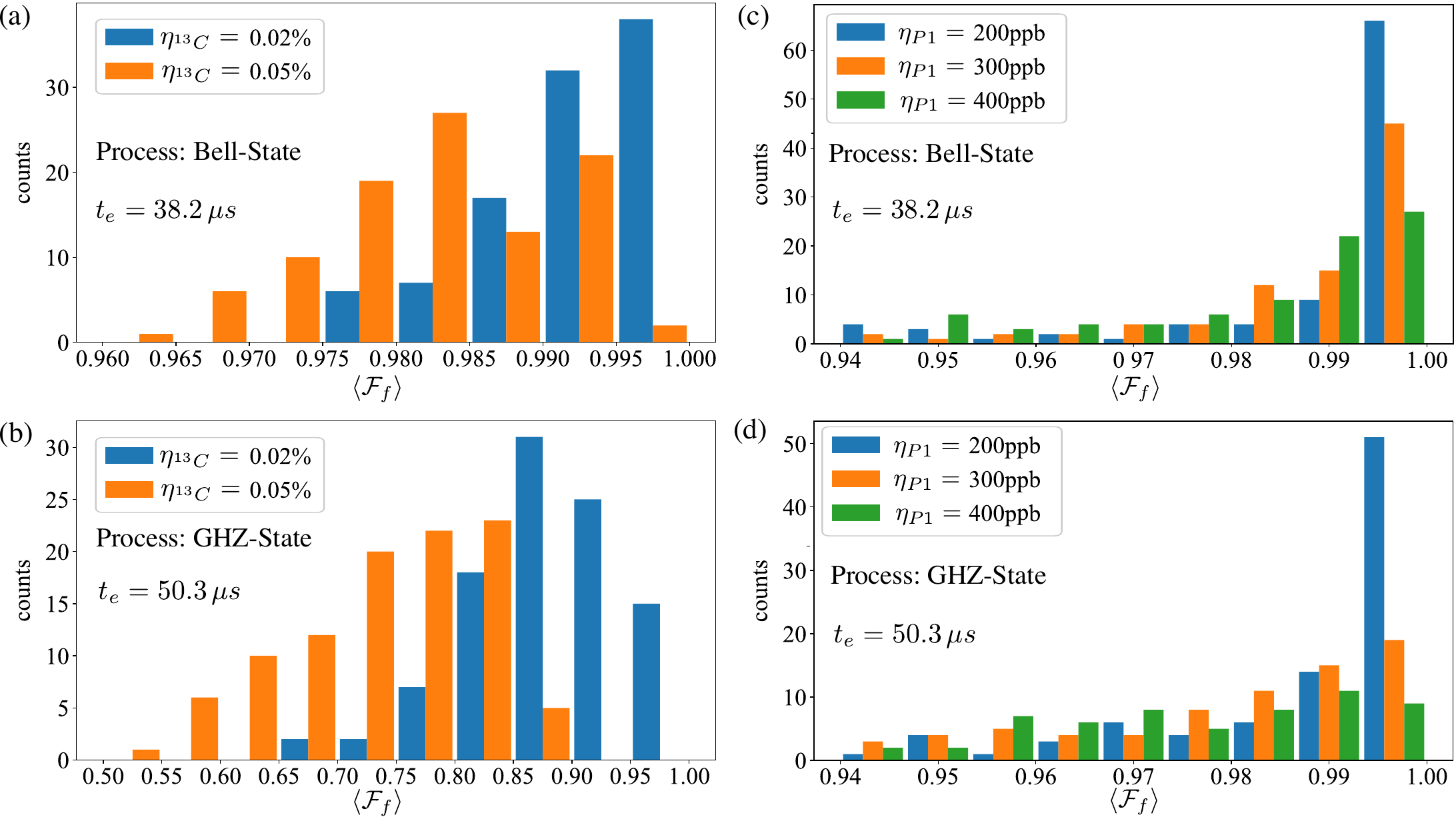}
%\scalebox{.8}{\input{Bild.tex}}
\caption{(a)-(b) Histograms for the fidelity of the Bell- and GHZ-process with a nuclear spin environment for two different abundances $\eta_{^{13}C}$. The histograms are calculated for 100 different spin-baths and sampled over 800 different bath states. (c)-(d) Analog histograms for the fidelity of the Bell- and GHZ-process in a P1-environment for three different abundances $\eta_{\text{P1}}$ calculated by sampling over 200 bath states (see also text).  } 
\label{Fig.7}
\end{figure*}

In summary, we can see that the dynamics of P1-spin-bath is not affected by the gate sequence and therefore its protection only depends on the amount decoupling pulses per time. On the other hand, this simple scheme fails for nuclear spin-baths, because the entanglement with residual environmental nuclear spins is not necessary overcome by the $\pi$-pulses but can even yield  more dynamics in the environment compared to the unprotected sequence.  

\subsection{Fidelities of multi-spin  quantum processes}

\subsubsection{Multi-spin processes}
In the following, we study the full system depicted in Fig.~\ref{Fig.1}a containing three close $^{13}C$ nuclear spins that, together with the NV-electron spin, form the register. To control this register, we apply 15 control pulses interleaved by 15 DD-pulses. We find appropriate parameters for the sequence by optimization to target states.
We study two different processes in this section. The first one is an entanglement gate between $^{13}C^{(1)}$ and $^{13}C^{(2)}$ leaving $^{13}C^{(3)}$ untouched. Further, the electron spin is supposed to be in the $m_s=0$ state at $t=t_e$, hence
\begin{align}
   \!|0,\frac{1}{2},\frac{1}{2},\frac{1}{2}\rangle \!\xrightarrow{\mathcal{S}^{\text{Bell}}_{\text{DD}}(\boldsymbol{t},\boldsymbol{\phi},\boldsymbol\theta) } \!  |0\rangle_{NV}\!\otimes\!|\text{Bell}\rangle_{^{13}C} \! \otimes\! |\frac{1}{2}\rangle_{^{13}C^{(3)}},
\end{align}
where $|\text{Bell}\rangle_{^{13}C}=\frac{1}{\sqrt{2}}\left(|\frac{1}{2},\frac{1}{2}\rangle+|-\frac{1}{2},-\frac{1}{2}\rangle\right)$.
We find a total evolution time $t_e=38.2\,\mu s$ with a process fidelity of $F_p\approx 0.9996$. The corresponding parameters for the gate sequence are given in the Appendix in Table~\ref{tab1}. 

The second target state we tend to achieve is a GHZ-state between all three $^{13}C$ spins with the electron spin again in the $m_s=0$ state after the sequence. 
The process can be expressed as
\begin{align}
   |0,\frac{1}{2},\frac{1}{2},\frac{1}{2}\rangle \xrightarrow{\mathcal{S}^{\text{GHZ}}_{\text{DD}}(\boldsymbol{t},\boldsymbol{\phi},\boldsymbol\theta)} |0\rangle_{NV}\otimes| \text{GHZ}\rangle,
\end{align}
where $|\text{GHZ}\rangle= \frac{1}{\sqrt{2}}\left(|\frac{1}{2},\frac{1}{2},\frac{1}{2}\rangle +|-\frac{1}{2},-\frac{1}{2},-\frac{1}{2}\rangle\right)$.
Since $^{13}C^{(3)}$ is farther away from the electron spin ($\sim1.5$ nm) the whole sequence takes longer ($t_e = 50.2\, \mu s$) and has a lower fidelity of $\mathcal{F}_p\approx 0.998$. We show the corresponding parameters also in Table~\ref{tab1} . 

\subsubsection{Fidelity histograms}
After having benchmarked the gCCE with a small spin system, we go beyond exactly treatable models and not only increase the number of spins in the register but also in the environment.
As explained in the methods section, the calculation of the gCCE approximation for the full density matrix for the four spin register becomes more challenging. 
Hence, here we restrict our attention to approximate the fidelities of the processes via the expansion. 
Particularly, we show histograms for the fidelity of the above mentioned quantum processes within the four-spin-register in Fig.~\ref{Fig.7}. Each histograms is calculated for $100$ different spin-bath configurations.  The register is either embedded in a nuclear spin-bath (a-b) or in a P1-spin-bath (c-d) with different abundances. 
Fig.~\ref{Fig.7}a shows the histogram of the Bell process for the two $^{13}C$ nuclear spins for abundances of $\eta_{^{13}C} \in \{ 0.02\,\%,0.05\,\%\}$ (sampled over $800$ bath states). We find that the probability of preparing the system with fidelity $\langle \mathcal{F}_f \rangle >0.99$ is still very high. Furthermore, spins that a further away than $4.0$ nm do not contribute to the gCCE1 contribution defining the \textit{impact radius} $r_c=4.0$ nm meaning that the system entangles with on average $7$ bath-spins for abundances $\eta_{^{13}C}=0.02\%$ and $19$ bath-spins for $\eta_{^{13}C}=0.05\%$. 

In Fig.~\ref{Fig.7}b, we show the histogram for the GHZ-state preparation and we find that the fidelities are strongly reduced. This can be attributed to the fact, that process takes longer yielding a larger \textit{impact radius} of the process. 
Namely, we find for the GHZ-gate that the results converged for a cutoff radius $r_c\approx 5.0$ nm, yielding to a gCCE1 interaction with on average $16$ bath-spins for $\eta_{^{13}C}=0.02\%$ and $42$ bath-spins for $\eta_{^{13}C}=0.05\%$. 
Hence, because we need to coherently control the nuclear spin at a distance at $1.5$ nm also the unwanted entanglement to more residual nuclear spins increases leading to a worse performance. 
Dividing the \textit{impact radius} $r_c$ throught the process time $t_e$, we find a \textit{butterfly velocity} of $\nu_B \approx 0.1 \frac{nm}{s}$ for both processes.

%%%%HERE%%%%%%
In Fig.~\ref{Fig.7}c-d we analogously show the performance of the processes within a P1-impurity spin-bath. The histograms are sampled over 200 different bath states. We find that results for $ \langle \mathcal{F}_f \rangle > 0.94$ are sufficiently converged with this sampling. Lower fidelities $\langle \mathcal{F}_f \rangle < 0.90$ would a need larger number of samples to keep the standard deviation below  $1\%$. We show the results (within gCCE2 approximation) for P1-spin-baths created within a spherical shell with $r_{min}=10$ nm and $r_{max}=100$ nm, for three different abundances and find that the interactions within the environment deteriorate the fidelity significantly. 
However, for the Bell process displayed in (c) most of the fidelities are still above $99\%$.
For the GHZ-process this is not the case and especially for $\eta_{P1}\gtrsim 300$ ppb, the probability of a high fidelity process ($\langle \mathcal{F}_f \rangle > 99\%$) is below $20\%$ (see Fig~\ref{Fig.7}d).
Since both processes have the same amount of decoupling pulses this strong change can be attributed to the fact that the process in (c) has more decoupling pulses per time ($0.39 \,\mu s^{-1}$) compared to (d) ($0.28 \,\mu s^{-1}$). 
This means that the Bell process is protected from more P1-P1 interactions than the GHZ process leading to higher fidelities. 

To systematically decouple from the correlations of such dense spin-baths, the evolution times between the decoupling pulses have to be smaller than the inverse of the strongest interaction between the spin-pair that is located closest to the system \cite{Bauch2020}.
Because for the spin-baths described in this section, close  electron spins in the environment interact with coupling strengths of the order of hundreds of MHz with the system \textit{and} with each other, one can estimate $\Delta t\ll 1 \,\mu$s to efficiently protect the system in spin-baths with $\eta_{P1}> 300$ ppb. On these timescales also the assumption of instantaneous pulses becomes questionable. 

Since, in real devices each decoupling pulse is not perfect, an optimal number of pulses to protect from such correlations within the environment has to be found to also perform long time calculations within a diamond based quantum register. 

\section{Discussion and Outlook}\label{Sec.Summary}
In summary, we theoretically studied the performance of a quantum register in diamond in presence of different spin-baths. 
The register is build of one $\text{NV}^{-}$ and three close by $^{13}C$ spins. 
We initiate the processes by only pulsing the NV electron spin and studied the corresponding dissipative time evolution of the full system. 
To model the spin-bath, we used the generalized cluster correlation expansion (\cite{Galli2021}) for the full register. 
We benchmarked this technique via an exact diagonalization of a small spin register within a spin-bath with nine impurities. 

The studied pulse sequences are very susceptible to surrounding nuclear spins. The main problem here is the accumulation of entanglement with unwanted residual nuclear spins that are too far away ($>2.0$ nm) to be efficiently controlled. 
This can be seen as the downside of the minimal approach of using only few pulses with high impact on the register. 
The impact radius of such a minimal sequence is difficult to control.
Since this effect is already captured by the gCCE1 contribution of the cluster expansion, the shown simulation scheme is very efficient. 
Further, we showed that in an environment consisting of P1-impurity spins the correlations within the spin-bath can become important within several tens of microseconds and the second order in the expansion (gCCE2) is sufficient on these timescales. 
For practical purposes, we found that only taking into account correlations between closest pairs of spins can already give the desired correction to the gCCE1 contribution. While keeping the accuracy of a full gCCE2-simulation, it substantially speeds up the computation.
For a four spin register, we analyzed two different quantum gates and their fidelities with experimentally realistic parameters and depending on the 
abundance of nuclear- or P1-impurities. This yields quantitative predictions how gate-performances deteriorate with increasing impurity densities and thus provides thresholds to be met by future material design and pulse shaping.
We find that for abundances of $\eta_{\text{P1}} \gtrapprox 200$ ppb the timescales of the spread of correlation within the environment yields errors even for DD pulse sequences. To efficiently decouple from theses effects a larger number or more sophisticated DD-pulses have to be used.
To avoid the detrimental effect of the nuclear spin-bath, the pulse sequences have to be designed more carefully. To make sure that unwanted spectator spins remain unaddressed, the processes are usually performed within magnetic fields $B_z>300$ G, where the Larmor frequency of the nuclear spins is much larger then the dipolar coupling to the electron spin. In this regime the number of pulses to control the register increases significantly leading to longer processes that are also stronger affected by control errors. Further, direct Rf-control of the nuclear spins also yields longer pulse times and makes processes slow. 
Nevertheless, we point out, that the minimalistic pulse sequence we study in this paper, performs well as along as not many $^{13}C$ spins have to be addressed.
Especially the fidelity for processes with abundances $\eta_{^{13}C} = 0.02\,\%$ are high (see Fig.~\ref{Fig.7}a,b) and also the possibility of finding such a register $P_R \approx 5 \, \%$ (see Fig.~\ref{Fig.2n}) is reasonable. Hence, there might be an optimal point regarding the abundances $\eta_{^{13}C}$, where the studied pulse sequences can be used for small registers in sparse nuclear spin environments. 
We propose, that the control scheme can be very useful in cases, where several small registers as depicted in Fig.~\ref{Fig.1}a are combined to build a larger register. 
Here, a suitable way to couple or arrange the individual NV-centers has to be found.

The methodology presented here allows to efficiently study different parameter regimes, sequences and error correction schemes for consecutive gate operations \cite{Suter2014,Plenio2015,Suter2015,Alber2005,Preskill2011,Lukin2020,Taminiau2021,Wrachtrup2014,Economou2023,Economou20232}. Our findings thus contribute to the rising activities in the less explored field of how to efficiently protect multi NV-based quantum gates from surrounding degrees of freedom. Since they are based on a framework treating the complete density matrix of the entire register, they offer a realistic test-bed for future concepts with quantitative predictions. 
\acknowledgments
We are grateful to R. Ghassemizadeh, W. Hahn, D. Hähnel, R. Said, J. Stockburger and B. Donvil
for fruitful discussions.
This work has been supported by the State of Baden - Württemberg under KQCBW/SiQuRe.

\subsection*{Data availability}
The data that support the findings of this study are available upon reasonable
request to the corresponding author.

\subsection*{Code availability}
The codes used in this study are available upon reasonable request to the
corresponding author.

\subsection*{Competing interests}
The Authors declare no Competing Financial or Non-Financial Interests.

\subsection*{Author contributions}
D.M. designed and wrote the numerical simulations. All authors analyzed the results, developed the project and contributed to writing the paper.

\appendix

\begin{table*}[t] 
\begin{tabular}{| c |c | c | c | c| c |c| c| c | c| c | c | c | c | c | c | c |}
\hline
 & 0 & 1  & 2 & 3 & 4 & 5 & 6 & 7 & 8 & 9 & 10 & 11 & 12 & 13 & 14 & 15\\
\hline
 {\textbf{two spins: Bell}} & & & & & & & & & & & & & & & &\\
\hline
  $\boldsymbol{t}[\mu\text{s}]$ & $0$ & $4.06$  &$1.57$ & $1.51$ & $-$ & $-$& $-$& $-$& $-$ & $-$ & $-$ & $-$ & $-$ & $-$ & $-$ & $-$  \\
 \hline
  $\boldsymbol{\theta}[\text{rad}]$ & $ - $ & $1.00$ & $3.58$ & $1.68$ & $-$ &$-$& $-$& $-$ & $-$ & $-$ & $-$ & $-$ & $-$ & $-$ & $-$& $-$ \\ 
 \hline
  $\boldsymbol{\phi}[\text{rad}]$ & $ - $  & $0.69$ & $1.97$ & $0.50$ & $-$ &$-$& $-$& $-$& $-$ & $-$ & $-$ & $-$ & $-$ & $-$ & $-$ & $-$ \\
  \hline
{\textbf{two spins: Bell-DD}} & & & & & & & & & & & & & & & &\\
\hline
  $\boldsymbol{t}[\mu\text{s}]$ & $0$ & $3.93$  &$2.95$ & $2.28$ & $0.30$ & $-$& $-$& $-$& $-$ & $-$ & $-$ & $-$ & $-$ & $-$ & $-$ & $-$  \\
 \hline
  $\boldsymbol{\theta}[\text{rad}]$ & $ - $ & $-1.06$ & $4.68$ & $ 2.07$ & $0$ &$-$& $-$& $-$& $-$ & $-$ & $-$ & $-$ & $-$ & $-$ & $-$ & $-$\\ 
 \hline
  $\boldsymbol{\phi}[\text{rad}]$ & $ - $  & $-0.53$ & $6.26$ & $-1.76$ & $0$ &$-$& $-$& $-$& $-$ & $-$ & $-$ & $-$ & $-$ & $-$& $-$ & $-$ \\
\hline
\hline
 {\textbf{GHZ-state $^{13}C$}} & & & & & & & & & & & & & & & &\\
\hline
  $\boldsymbol{t}[\mu\text{s}]$ & $ 1.32 $ & $4.26$  &$7.01$ & $5.42$ & $7.15$ & $5.21$ & $2.83$ & $2.68$ & $2.32$ & $4.05$ & $1.73$ & $3.85$ & $0.19$ & $0.08$ & $1.82$ & $0.32$ \\
 \hline
  $\boldsymbol{\theta}[\text{rad}]$ & $-$ & $1.74$  &$-0.76$ & $4.01$ & $2.48$ & $3.61$ & $6.70$ & $-1.11$ & $5.97$ & $1.65$ & $1.26$ & $-1.35$ & $5.74$ & $3.13$ & $2.89$ & $0.35$  \\ 
 \hline
  $\boldsymbol{\phi}[\text{rad}]$ & $-$ & $5.97$ & $5.58$  &$-3.53$ & $6.88$ & $4.43$ & $-1.43$ & $-0.84$ & $5.47$  & $2.30$ & $0.10$ & $0.85$ & $3.03$ & $2.68$ & $1.64$ & $1.44$  \\
  \hline
  \hline
 {\textbf{Bell-state $^{13}C$}} & & & & & & & & & & & & & & & &\\
\hline
  $\boldsymbol{t}[\mu\text{s}]$ & $1.23$ & $1.55$  &$3.26$ & $2.90$ & $0.057$ & $2.93$ & $2.98$ & $2.69$ & $3.40$  & $4.10$ & $0.87$ & $3.47$ & $2.25$ & $1.02$ & $2.03$ & $3.44$ \\
 \hline
  $\boldsymbol{\theta}[\text{rad}]$ & $-$ & $1.44$  &$-0.79$ & $4.23$ & $-0.37$ & $2.23$ & $4.58$ & $1.04$ & $3.07$ & $1.05$ & $1.03$ & $-2.08$ & $3.61$ & $4.75$ & $2.34$ & $1.88$  \\
 \hline
  $\boldsymbol{\phi}[\text{rad}]$ & $-$ & $5.92$  &$8.07$ & $-2.86$ & $ 7.02$ & $3.96$ & $0.24$ & $-0.70$ & $7.10$ & $1.14$ & $0.15$ & $2.21$ & $1.82$ & $3.04$ & $1.83$ & $0.33$ \\
  \hline
 % \label{tab.1}
\end{tabular}

 \caption{Parameters for the pulse sequences for the processes simulated in the main text as described in Fig.~\ref{Fig.1}b and Fig.~\ref{Fig.pulsesdd} }
 \label{tab1}
\end{table*}

\section{Dipole interaction tensors}\label{App.A}
To make the specific nature of the simulated register more compareble to experiments, we show the dipole interaction tensors of the fully connected register in this Appendix. We start with $\bar{M}$ encoding the interaction between the $\text{NV}^{-}$ and the $^{13}C$ nuclear spins 
\begin{align}
\!\!\boldsymbol{S}\bar{M}^{(k)}\boldsymbol{I}^{(k)}_C\!=\!-\frac{\hbar\mu_{0}}{4 \pi} \frac{\gamma_{e} \gamma_{C} }{r^{3}}[3(\boldsymbol{S}\!\cdot\!\hat{n})(\boldsymbol{I}^{(k)}_{C}\! \cdot \! \hat{n})\!-\!\boldsymbol{S}\! \cdot \!\boldsymbol{I}^{(k)}_C],
\end{align}
where $r$ is the distance of the $^{13}C$ to the $\text{NV}^{-}$ electron spin, $\hat{n}$ the unit vector pointing to the $^{13}C$, $\gamma_e=-1.761\times10^{11}\,T^{-1}s^{-1}$ and $\gamma_C=6.728\times10^7\,T^{-1}s^{-1}$ are the gyromagnetic ratios of the electron and the nuclear spin, respectively. 
The first $^{13}C$ (k=1) has the spherical coordinates $\boldsymbol{r} = (r = 0.89\,\text{nm},\varphi = 0^\circ, \vartheta  = 78^\circ )$ yielding
\begin{align}
   \bar{M}^{(1)}_{xx}& \approx 328.8 \text{kHz} ,\\  \nonumber    \bar{M}^{(1)}_{y y} &\approx -175.8  \text{kHz},\\   \nonumber    \bar{M}^{(1)}_{zz}& \approx -153.0 \text{kHz},\\ \nonumber  
   \bar{M}^{(1)}_{xy} & =    \bar{M}^{(1)}_{yx}=    \bar{M}^{(1)}_{yz}=   \bar{M}^{(1)}_{zy}=0, \\ \nonumber  
   \bar{M}^{(1)}_{zx} &=    \bar{M}^{(1)}_{xz}  \approx 107.3\text{kHz}. 
\end{align}
Analogously, we find for the second $^{13}C$ (k=2) having the spherical coordinates $\boldsymbol{r} = (r = 1.0\text{nm},\varphi = 53.97^\circ, \vartheta  = 107.94^\circ )$ 
\begin{align}
   \bar{M}^{(2)}_{xx}& \approx -7.809 \text{kHz} ,\\  \nonumber    \bar{M}^{(2)}_{y y} &\approx 96.96  \text{kHz},\\   \nonumber    \bar{M}^{(2)}_{zz}& \approx -89.15 \text{kHz},\\ \nonumber  
   \bar{M}^{(2)}_{xy} & =    \bar{M}^{(2)}_{yx}\approx 161.7 \text{kHz}  ,\\ \nonumber
   \bar{M}^{(2)}_{yz} &=   \bar{M}^{(2)}_{zy}\approx -64.75\text{kHz}, \\ \nonumber  
   \bar{M}^{(2)}_{zx} &=    \bar{M}^{(2)}_{xz}  \approx -89.12 \text{kHz}. 
\end{align}
For the last $^{13}C$ (k=3) having the spherical coordinates $\boldsymbol{r} = (r = 1.5\text{nm},\varphi = -71.9^\circ, \vartheta  = -53.97^\circ )$, we find
\begin{align}
   \bar{M}^{(3)}_{xx}& \approx -2.313 \text{kHz} ,\\  \nonumber    \bar{M}^{(3)}_{y y} &\approx 28.72  \text{kHz},\\   \nonumber    \bar{M}^{(3)}_{zz}& \approx -26.41 \text{kHz},\\ \nonumber  
   \bar{M}^{(3)}_{xy} & =    \bar{M}^{(2)}_{yx}\approx -47.77 \text{kHz}  ,\\ \nonumber
   \bar{M}^{(3)}_{yz} &=   \bar{M}^{(3)}_{zy}\approx -19.19\text{kHz}, \\ \nonumber  
   \bar{M}^{(3)}_{zx} &=    \bar{M}^{(3)}_{xz}  \approx -26.40 \text{kHz}. 
\end{align}
The interaction tensors between the two $^{13}C$ and between the $^{13}C$s and the nitrogen nuclear spin do not play a role in this specific register since the interactions are of the order of several Hz. However, we included it into our simulation for future analysis. 
Since the interaction between the nuclear spin nitrogen atom and the electron spin of the NV is not solely defined by the dipolar interaction but also by the Fermi-Contact contribution, we use $N_{\parallel}=-1.73\,$MHz and $N_{\perp}=-2.16\,$MHz as given in \cite{Korener2021}. 
%Since the couplings to the bath spins varies for each configuration, we cannot give any specific values. 
%
%
\begin{figure}[t]
\includegraphics[scale=0.52]{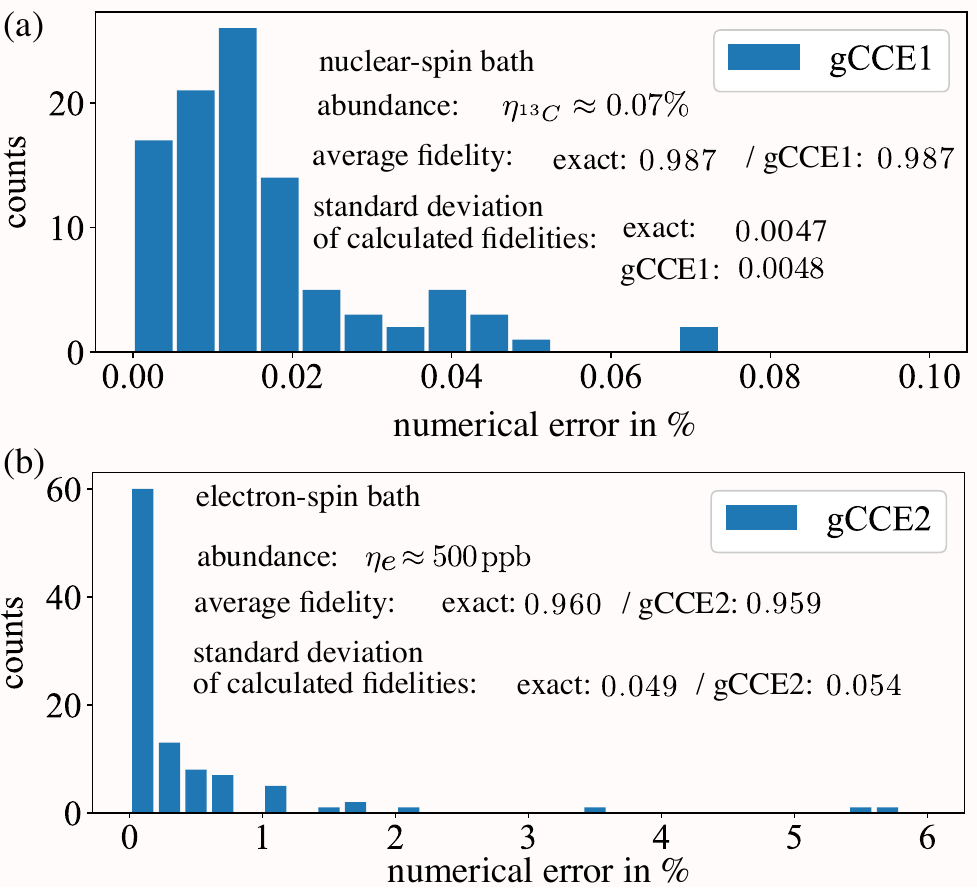}
%\scalebox{.8}{\input{Bild.tex}}
\caption{Histograms for the relative numerical error of the gCCE1/gCCE2, respectively - calculated directly on the fidelity - with respect to the exact diagonalization, for the of two-spin entanglement process (see Sec.\ref{Sec.twospins}). (a) Nuclear spin-bath consisting of nine spins with abundance $\eta_{^{13}C} \approx 0.07\%$. The numerical error remains below $0.1\%$ for all the simulated fidelities. (b) The spin-bath consists nine pure electron spins (not P1-centers) which are strongly interacting. The abundance is $\eta_e\approx 500$ ppb. We find that most of the results have numerical errors below $0.2\%$. The outliers can be attributed to the necessity of calculating also the gCCE3 contribution due to the strong interaction between bath-electrons. We attribute the errors $\lesssim 1\%$ to small errors in the calculation of each of the elements in the products of Eq.~(\ref{Eq.B}) which accumulate (i.e. the errors in (a)). In both histograms, the gCCE was sampled over all possible bath states. Benchmarking of further statistical quantities of the produced fidelity-data of the 100 different spin-baths, is done within the figures. } 
\label{App1}
\end{figure}

\section{gCCE: Higher orders and Fidelity}\label{App.B}
Here we briefly dicuss the calculation of higher orders in the gCCE. For the gCCE2 contribution, we start with the density matrix 
\begin{align}
    \tilde{\rho}^{(n,l,q)}_{\text{E2}}(0) = \rho(0) \otimes \rho^{(n)}_{l,q}(0), 
\end{align}
where we included two spins ($l$ and $q$) of the environment into the full density matrix. The reduced density matrix then becomes
\begin{align}
        \rho^{(n,l,q)}_{\text{E2}} (t_e)= \text{tr}_{B,l,q}\left(\mathcal{S}_{n,l,q}\,    \tilde{\rho}^{(n,l,q)}_{\text{E2}}(0)\, \mathcal{S}_{n,l,q}^\dagger \right).
\end{align}
To obtain the correction to the gCCE1, we use
\begin{align} 
\langle a | \rho_{\text{E2}}^{(n)}(t)| b\rangle &=  \langle a | \rho^{(n)}_{\text{E0}}(t)| b\rangle\, \left[\prod_l  \frac{\langle a|\rho^{(n,l)}_{\text{E1}}(t)|b\rangle}{\langle a | \rho^{(n)}_{\text{E0}}(t)| b\rangle}\right]\label{Eq.gCCE2} \\& \!\!\!\!\!\!\!\!\times \prod_{l, q} \frac{\langle a|\rho^{(n,l,q)}_{\text{E2}}(t)| b\rangle}{\langle a | \rho^{(n)}_{\text{E0}}(t)| b\rangle^{-1}\langle a|\rho^{(n,l)}_{\text{E1}}(t)|b\rangle\langle a|\rho^{(n,q)}_{\text{E1}}(t)|b\rangle}. \nonumber
\end{align}
Averaging over different bath-states $n$ yields the gCCE density matrix to second order. Generally j-th order correction of the gCCE can be recursively written as
\begin{align}
\langle a |\tilde{\rho}^{(n)}_{Ej}|b\rangle=\frac{\left\langle a\left|\rho_{Ej}^{(C_j)}\right| b\right\rangle}{\prod_{{C}^{\prime}} \langle a|\tilde{\rho}_{\left\{{C}^{\prime} \subset C_j\right\}}|b\rangle},
\end{align} 
where $C_j$ contains a number of indices coinciding with number of  included bath spins $j$ (e.g. $C_2\rightarrow l,p$ and $C_3\rightarrow l,p,s$). Using this definition a general order $j$ reads 
\begin{align}
\langle a | \rho_{\text{Ej}}^{(n)}| b\rangle &= \langle a | \rho^{(n)}_{\text{E0}}| b\rangle\, \left[\prod_{C_1} \langle a |\tilde{\rho}^{(n)}_{E1}|b\rangle\right]\\&\times \left[\prod_{C_2} \langle a |\tilde{\rho}^{(n)}_{E2}|b\rangle\right] ...\times\left[\prod_{C_j} \langle a |\tilde{\rho}^{(n)}_{Ej}|b\rangle\right]\nonumber. 
\end{align}
This result is again to be averaged over different states $n$. For more information we refer to \cite{Galli2021}.
%%%%%%%%%%%%%%%%%%%%%%%%%%%%%%%%
%\begin{figure}[t]
%\includegraphics[scale=0.8]{Fig9.pdf}\\
%\scalebox{.8}{\input{Bild.tex}}
%\caption{$S_z$ component of the NV electron spin under the influence of the pulse sequences for the different quantum processes on the full register.  }
%\label{Fig.4}
%\end{figure}
%%%%%%%%%%%%%%%%%%%%%%%%%%%%%%%%

Furthermore, due to numerical stability arguments the histogram Fig.~\ref{Fig.7} displaying the gCCE approximation for the four qubit register, was calculated by using the expansion directly for the fidelity, i.e. 
\begin{align}
\mathcal{F}^{(n)}_{f,\text{E2}} &= \mathcal{F}^{(n)}_{f,\text{E0}} \left[\prod_l  \frac{\mathcal{F}^{(n,l)}_{f,\text{E1}}}{\mathcal{F}^{(n)}_{f,\text{E0}}}\right]  \prod_{l, q} \frac{\mathcal{F}^{(n,l,q)}_{f,\text{E2}}}{(\mathcal{F}^{(n)}_{f,\text{E0}})^{-1}\mathcal{F}^{(n,l)}_{f,\text{E1}}\mathcal{F}^{(n,q)}_{f,\text{E1}}}. \label{Eq.B}
\end{align}
We checked the equivalence to using Eq.~(\ref{Eq.gCCE2}) together with Eq.~(\ref{Eq.firstfid}) numerically for the two spin register and with respect to the exact diagonalization (see also Figs.~\ref{App1}). Furthermore, it is easy to show that
\begin{align}
\mathcal{F}^{(n,l,q)}_f\!=\!\frac{\text{tr}\left(\rho_T^\dagger \rho^{(n,l,q)}(t_e)\right)}{\text{tr}\left(\rho_T^\dagger \rho_T\right)}=\frac{\langle 0|\langle n|\bar{\rho}^{(n,l,q)}_T(-t_e)|n\rangle|0\rangle}{\text{tr}\left(\rho_T^\dagger \rho_T\right)}, \label{Eq.newgCCE}
\end{align}
where $|0\rangle = |0\rangle_{NV}\otimes |1/2\rangle_{^{13}C}  \otimes  |1/2\rangle_{^{13}C} ...
$ corresponds to the initial state of the register and $|n\rangle$ to the initial state of the bath,
\begin{align}
 \bar{\rho}^{(n,l,q)}_T(-t_e) = \mathcal{S}_{n,l,q}^\dagger\, (\rho^\dagger_T \otimes \rho^{(l,q)}_B)\, \mathcal{S}_{n,l,q}
\end{align}
and $\rho^{(l,q)}_B$ is the completely mixed density matrix of the spins $l$ and $q$ in the environment. Hence, we can interpret the fidelity as a back evolution of the density matrix $(\rho^\dagger_T \otimes \rho^{(l,q)}_B)$. Since the fidelity can be written as a single time evolved density matrix element, we can use the gCCE directly on $\mathcal{F}^{(n)}_f$. Please note that this argument only holds for pure initial states of the system density matrix (here $|0\rangle$). 
\section{Table of pulse sequences}\label{App.C}
We display the parameters for the microwave pulse sequences (as explained in  Fig.~\ref{Fig.1}b and Fig.~\ref{Fig.pulsesdd}) used to initiate the quantum processes on the register in table~\ref{tab1}. Further, we visualize the $\langle S_z \rangle$ component for the processes on the four spin register (without DD) in Fig.~\ref{Fig.pulsesdd}.

\bibliographystyle{mybibstyle}
 
\bibliography{bibliography}

\end{document}